\documentclass[11pt]{article}
\usepackage{booktabs} 
\usepackage[ruled]{algorithm2e} 

\SetAlFnt{\small}
\SetAlCapFnt{\small}
\SetAlCapNameFnt{\small}
\SetAlCapHSkip{0pt}
\usepackage[utf8]{inputenc}
\usepackage{fullpage}
\usepackage{amsmath,amsfonts,amsthm}
\usepackage{thmtools} 
\usepackage{xcolor}
\usepackage{amssymb}
\usepackage{bm}
\usepackage{xspace}
\usepackage{mathtools}
\usepackage{hyperref}
\usepackage{nicefrac}
\usepackage[numbers]{natbib}

\usepackage{comment}

\newtheorem{theorem}{Theorem}
\newtheorem{lemma}{Lemma}
\newtheorem{proposition}{Proposition}
\newtheorem{claim}{Claim}
\newtheorem{corollary}{Corollary}
\newtheorem{definition}{Definition}

\newtheorem{fact}{Fact}

\newtheorem{assumption}{Assumption}

\DeclareMathOperator*{\argmax}{arg\,max}
\DeclareMathOperator*{\argmin}{arg\,min}

\def\R{\mathbb{R}}

\newcommand{\E}{\mathbb{E}}

\renewcommand{\gg}{\bm{\gamma}}
\renewcommand{\k}{\hat{k}}

\newcommand{\tcpa}{{ \mbox{tCPA} }}
\newcommand{\ql}{{ \mbox{QL} }}
\newcommand{\budgeted}{{ \mbox{Budgeted} }}
\newcommand{\budget}{{ \mbox{B} }}
\newcommand{\PPAD}{{\texttt{PPAD}}}

\def\b{{\bm{b}}}
\def\r{{\bm{r}}}
\def \x {{\bm{x}}}

\def\x{{\bm{x}}}

\def \cC {\mathcal{C}}
\def \cD {\mathcal{D}}
\def \cG {\mathcal{G}}
\def \cI {\mathcal{I}}
\def \cT {\mathcal{T}}
\def \eps {\varepsilon}

\ifdefined\NOCOMMENT
    \newcommand{\ganote}[1]{}
    \newcommand{\apnote}[1]{}
    \newcommand{\jznote}[1]{}
    
\else
    \newcounter{note}[section]
    \renewcommand{\thenote}{\thesection.\arabic{note}}

\newcommand{\apnote}[1]{\refstepcounter{note}\textcolor{blue}{$\ll${\bf Andres~\thenote:} {\sf #1}$\ggg$\marginpar{\tiny\bf AP~\thenote}}}
    
\newcommand{\ganote}[1]{\refstepcounter{note}\textcolor{blue}{$\ll${\bf Gagan~\thenote:} {\sf #1}$\ggg$\marginpar{\tiny\bf GA~\thenote}}}
    
\newcommand{\jznote}[1]{\refstepcounter{note}\textcolor{brown}{$\ll${\bf Junyao~\thenote:} {\sf #1}$\ggg$\marginpar{\tiny\bf JZ~\thenote}}}

\setcitestyle{acmnumeric}

\title{Multi-Channel Auction Design in the Autobidding World}

\author{Gagan Aggarwal\thanks{Google, \texttt{\{gagana,perlroth\}@google.com}}, Andres Perlroth\footnotemark[1]\; and Junyao Zhao\thanks{Stanford University, \texttt{junyaoz@stanford.edu}}}

\usepackage{color}

\begin{document}

\begin{titlepage}

\maketitle
\begin{abstract}
Over the past few years, more and more Internet advertisers have started using automated bidding for optimizing their advertising campaigns. Such advertisers have an optimization goal (e.g. to maximize conversions), and some constraints (e.g. a budget or an upper bound on average cost per conversion), and the automated bidding system optimizes their auction bids on their behalf. Often, these advertisers participate on multiple advertising channels and try to optimize across these channels. A central question that remains unexplored is how automated bidding affects optimal auction design in the {\em multi-channel setting}. 

In this paper, we study the problem of setting auction reserve prices in the multi-channel setting. In particular, we shed light on the 
revenue implications of whether each channel optimizes its reserve price {\em locally}, or whether the channels optimize them {\em globally} to maximize total revenue.
Motivated by practice, we consider two models: one in which the channels have full freedom to set reserve prices, and another in which the channels have to respect floor prices set by the publisher. We show that in the first model, welfare and revenue loss from {\em local} optimization is bounded by a function of the advertisers' inputs, but is independent of the number of channels and bidders. In stark contrast, we show that the revenue from {\em local} optimization could be arbitrarily smaller than those from {\em global} optimization in the second model.
\end{abstract}

\end{titlepage}


\section{Introduction}

Advertisers are increasingly using automated bidding in order to set bids for ad auctions in online advertising. Automated bidding simplifies the bidding process for advertisers -- it allows an advertiser to specify a high-level goal and one or more constraints, and optimizes their auction bids on their behalf~\cite{googlebiddingpage,metabiddingpage,tiktokbiddingpage,twitterbiddingpage}. A common goal is to maximize conversions or conversion value. Some common constraints include Budgets and TargetCPA (i.e. an upper bound on average cost per conversion). 
This trend has led to interesting new questions on auction design in the presence of automated bidders~\cite{AggarwalBM19,DengMMZ21,BalseiroDMMZ21,balseiro2021robust}. 

One central question that remains unexplored is how automated bidding affects optimal auction design in the {\em multi-channel setting}. 
It is common for advertisers to show ads on multiple channels and optimize across channels. 
For example, an advertiser can optimize across Google Ads inventory (YouTube, Display, Search, Discover, Gmail, and Maps) with Performance Ads~\cite{googlemaxperformance}, or can optimize across Facebook, Instagram and Messenger with Automated Ad Placement~\cite{metaadplacement}, or an app developer can advertise across Google's properties including Search, Google Play and YouTube with App campaigns~\cite{googleuniversalappcampaigns}. With traditional quasi-linear bidders, the problem of auction design on each channel is independent of other channels' designs. However, when advertisers use automated bidders and optimize across channels, the auction design of one channel creates externalities for the other channels through the constraints of automated bidders.

Motivated by this, we introduce the problem of auction design in the multi-channel setting with automated bidding across channels. In particular, we study the problem of setting reserve prices across channels. We consider two behavior models: {\em Local} and {\em Global}. In the {\em Local} model, each channel optimizes its reserve price to maximize its own revenue, while in the {\em Global} model, the channels optimize their reserve prices {\em globally} in order to maximize the total revenue across channels. The main question is: {\em what is the revenue loss from optimizing {\em locally}, rather than {\em globally}}? 

We consider this question in two settings: one in which each channel has full control over its reserve prices, and one in which the channels have to respect an externally-imposed lower bound on the reserve prices. The first setting which we call {\em Without Publisher Reserves} is very common in practice and arises when the impressions are owned by the selling channel, or when the publisher leaves the pricing decisions to the selling channel. The second setting which we call {\em With Publisher Reserves} arises when the impressions are owned by a third-party publisher that sets a floor price for its impressions -- this could come from an outside option for selling the impression. This is common in Display advertising where the selling channel is often different from the publisher who owns the impressions.

{\bf Model:}
Our model consists of $k$ channels, each selling a set of impressions. Each channel can set a uniform reserve price. The uniform reserve price is in the cost-per-unit-value space\footnote{See Section~\ref{sec: further-discussions} for a discussion of uniform reserve prices in the cost-per-impression space} (see Section~\ref{subsec:bidders} for details). This is motivated by the observation that, in practice, values are commonly known by the channels; values are usually click-through-rate or conversion-rate of an ad, as in~\cite{AggarwalBM19}, and the channels have good estimates for those. Besides the reserve prices set by channels, in the {\em With Publisher Reserves} setting, each impression could have a price floor set by the publisher who owns the impression.
Each impression is sold in a Second-Price-Auction with a floor price that depends on the reserve price set by the selling channel and the price constraint set by the publisher. Bidders want to maximize their conversions (or some other form of value) subject to one of two types of constraints: (1) Budget, an upper bound on spend and (2) TargetCPA, an upper bound on the average cost per conversion. The model also allows standard quasi-linear bidders with no constraints.
The game consists of two main stages: First, each channel simultaneously announces its reserve price; then, bidders bid optimally for the different impressions. 

\subsection{Our results}
The paper's main focus is to compare the revenue\footnote{See Section~\ref{sec: further-discussions} for a brief discussion of {\em welfare}} at equilibrium when channels optimize locally, i.e. each sets its reserve price(s) to maximize its own revenue, to the revenue where channels act globally and set their reserve prices to maximize the sum of the total revenue.
We define the Price of Anarchy (PoA) as the worst-case ratio between the total revenue when the channels optimize locally compared to the case where the channels optimize globally. Our main goal is to bound the Price of Anarchy in the two settings: {\em Without Publisher Reserves}, and {\em With Publisher Reserves}.

\subsection*{Setting without Publisher Reserves}
In order to bound the Price of Anarchy, we first bound the local and global revenue in terms of the optimal Liquid Welfare 
(see Section~\ref{sec:model} for the definition). These revenue bounds are interesting in their own right and the proof methodology gives (non-polytime) algorithms for determining good reserve prices. 

We first consider the worst-case revenue in the local model where each channel is optimizing for its own revenue, compared to the optimal Liquid Welfare. We show in Theorem~\ref{thm:rev-local} that the worst-case revenue is at least $\Omega(\frac{1}{\log \eta})$ 
fraction of the optimal Liquid Welfare, where $\eta$ depends on the bidders' inputs\footnote{$\eta$ is the maximum of the ratio of the highest to lowest TargetCPA among TargetCPA bidders and a ratio defined (in Definition~\ref{def:budget-fraction}) for Budgeted bidders.} and quantifies the heterogeneity of the pool of bidders. This lower bound on revenue trivially carries over to the setting where the channels are optimizing globally and to the single-channel setting. Next, we show that this bound is tight up to constant factors (Proposition~\ref{ex:revg-ub}). In particular, we give an example in the single-channel setting where the optimal revenue with a uniform reserve price is $O(\frac{1}{\log \eta})$ of the optimal Liquid Welfare. This upper bound also applies to the global and local models in the multi-channel setting. In other words, the upper bound and lower bounds on the gap between Liquid Welfare and revenue in each of these settings is $\Theta(\log \eta)$. That naturally makes one wonder: Is optimizing locally as good for revenue as optimizing globally? If we look into the gap bounds, we find that they arise from trying to capture values of different scales with a uniform reserve price. And one might conjecture that since the source of the gap applies to both the global and local model, that even if there is a revenue gap between the two models, it should depend on different factors. Surprisingly, we show that the gap between optimizing locally and globally is exactly the same $\log \eta$ factor (Theorem~\ref{lem:tight-PoA}).
Note that in all the above settings, the revenue guarantee is independent of the number of channels and bidders and depends only on the heterogeneity of the bidders.

\subsection*{Setting with Publisher Reserves}
In stark contrast to the setting without publisher reserves, we show that the PoA in this setting can be arbitrarily small even with one $\tcpa$ bidder (Theorem~\ref{th:general-publisher-model}). The gap example depends heavily on the asymmetry between the different channels. Motivated by this, we consider the restricted setting where each channel sells a random sample of the impressions (see Section~\ref{publisher-reserve} for the exact details). For this case, with one $\tcpa$ bidder in the game, we show that under some mild constraint on the channels' strategies, the PoA = $1/k$, where $k$ is the number of channels. 
When the channels optimize globally, the equilibrium is efficient and all channels set low reserve prices. On the other hand, for the equilibrium in the local optimization model, the larger channels (in terms of the volume of impressions they own) set low reserve prices while small channels are extractive and set high reserve prices.

\subsection*{Hardness of Equilibrium Computation}
To complement our Price of Anarchy results, we also study
the computational complexity of computing the equilibrium of the game. 
We show an impossibility result -- that it is $\PPAD$-hard to compute the subgame equilibrium of the bidders (Theorem~\ref{thm:subgame-hardness}). To prove this result, we use gadget reduction from the problem of finding approximate Nash equilibrium for $0$-$1$ bimatrix game, and we need to handle many difficulties unique to our subgame, which we explain with more details in Section~\ref{sec:hardness} and Appendix~\ref{app:proof-of-hardness}.

\subsection*{Key implications of our results}
Our results have several implications for setting reserve prices in the multi-channel setting: 
\begin{itemize}
    \item The revenue gap between local and global optimization depends heavily on whether there are publisher-imposed reserve prices. 
    \item Without publisher reserves, the worst-case gap between the revenue in the local model and the global model is $\Theta(\log \eta)$, where $\eta$ captures the heterogeneity of the bidders' inputs and is independent of the number of channels and bidders. Thus it is better to optimize globally when possible. 
    \item Without publisher reserves, it is possible to obtain a revenue of $\Theta(\frac{1}{\log \eta})$ fraction of the optimal Liquid Welfare by setting uniform reserve prices. This observation is not surprising in the single-channel setting and for global optimization in the multi-channel setting, but it is remarkable that it holds even with local optimization, where the selfishness of a channel could have made it difficult for other channels to make revenue. We also note that the approximation can be improved by setting reserve prices at a more granular level, rather than a uniform reserve price. In that case, the approximation ratio will depend on the heterogeneity of bidders per slice.
    \item With publisher reserves, the gap between the revenue in the local and global model can be arbitrarily large. This can happen even when only one of the channels has external pricing constraint.
\end{itemize}

\subsection*{Organization of the paper}
We present a formal model of the problem in Section~\ref{sec:model}. Then, in Section~\ref{sec:hardness}, we show that it is $\PPAD$-hard to compute the equilibrium of the sub-game. In Section~\ref{zero-publisher-reserve}, we study the setting without publisher reserves and present a tight bound on the Price of Anarchy, as well as on the gap between the revenue and optimal Liquid Welfare in the local and global models. In Section~\ref{publisher-reserve}, we study the setting with publisher reserves and show the Price of Anarchy is 0. We also study a restricted version of this setting, and show a Price of Anarchy of $1/k$ for that version. Finally, in Section~\ref{sec: further-discussions}, we discuss extensions for welfare and for setting reserve prices in the cost-per-impression space.

\section{Related Work}
\medskip
\noindent {\bf Autobidding.} There has been a lot of recent interest in exploring questions related to automated bidding, including bidding algorithms and their equilibria~\cite{AggarwalBM19}, and auction design in the presence of automated bidding~\cite{DengMMZ21,BalseiroDMMZ21,balseiro2021robust}. \citet{AggarwalBM19} initiate the study of autobidding and find optimal bidding strategies for a general class of autobidding constraints. They also prove the existence of an equilibrium and prove a lower bound on liquid welfare at equilibrium compared to the optimal liquid welfare. \citet{DengMMZ21} show how boosts can be used to improve welfare guarantees when bidders can have both TargetCPA and Budget constraints, potentially at the cost of revenue. \citet{BalseiroDMMZ21} characterize the revenue-optimal single-stage auctions with either value-maximizers or utility-maximizers with TargetCPA constraints, when either the values and/or the targets are private. Similar to our paper, \citet{balseiro2021robust} also study reserve prices in the presence of autobidders, and show that with TargetCPA and Quasi-linear bidders, revenue and welfare can be increased by using (bidder-specific) reserve prices. They do not study budget-constrained bidders.
All of the above papers are in the single channel setting. \medskip

\noindent {\bf Auction design with multiple channels.}
Most of this stream of literature have focused on models where multiple channels (auctioneers) compete to take captive profit-maximizers buyers \cite{Burguet1999ImperfectCI, ellison, preston}. The competition across channels leads to lower reserve prices, obtaining lower revenues and more efficient outcomes~\cite{virag}. Our model differs from them in that our bidders are not captive but are instead are optimizing under their autobidding constraints. Interestingly, we show that in some cases the competition among channels leads to higher reserve prices, and at the same time, improves welfare (see Theorem~\ref{th: scaled channels PoA=0}.).

\section{Model}
\label{sec:model}
Our baseline model considers a set of bidders (advertisers) $J$ interested in purchasing a set of impressions $I$ that are sold by $K$ different channels.
The impressions that channel $k$ sells, $i\in I_k$, are sold using a second-price auction with a floor price. This floor price depends on the reserve price $r_k$ chosen by the channel and by the minimum price $p_{i}$ set by the publisher that owns the impression\footnote{The publisher might have an outside option to sell some of the impressions and sets a reserve price to account for that. These reserve prices are prechosen by the publishers and hence are fixed constants known to both channels and bidders.}.

\subsection{Bidders}
\label{subsec:bidders}
Motivated by the most common bidding formats that are used in practice, we assume that each bidder can be one of the following types: a {\em $\tcpa$} bidder, a {\em $\budgeted$} bidder, or a {\em Quasi-linear ($\ql$)} bidder. We denote by $J_{\tiny \tcpa}$, $J_{\tiny \budgeted}$ and $J_{\tiny \ql}$ the set of bidders that are $\tcpa$, $\budgeted$ and $\ql$ bidders, respectively.
 
Each Bidder $j$ has a value (e.g. conversion rate) $v_{j,i}$ for impression $i$ and submits a bid $b_{j,i}$ for the impression. A bidder's cost for buying impression $i\in I_k$ is 
\begin{align}
    & c_{j,i}(\b_{i},\r) = \max\{\max_{\ell\neq j\textrm{ s.t. } b_{\ell,i}\ge\max\{r_k v_{\ell,i},\,p_{i} v_{\ell,i}\}}\{ b_{\ell,i}\},\,r_k v_{j,i},\,p_{i} v_{j,i}\} \label{con:c},
\end{align}
where $\b_{i} = (b_{j,i})_{j\in J}$ (note we use the notation $c_{j,i}(\b_{i},\r)$ for simplicity, even though $c_{j,i}(\b_{i},\r)$ does not depend on $b_{j,i}$) and $\r=(r_k)_{k\in K}$ (because $p_i$'s are fixed constants prechosen by the publishers, for simplicity we do not include them as variables). That is, a bidder's cost for an impression is the maximum among (i) the bids of the bidders who bid above their own reserve prices, (ii) reserve price set by the channel which owns the impression, and (iii) reserve price set by the publisher. Also, note that the reserve prices $r_k$ and $p_i$ are multiplied by $v_{j,i}$ to get the final floor price of impression $i$ for bidder $j$. In other words, the reserve prices are in the cost-per-unit-value space. We will refer to the final reserve price of impression $i$ for bidder $j$ by $r_{j,i}:=\max\{r_k v_{j,i},\,p_{i} v_{j,i}\}$. Now we explain the bidder types.

\smallskip
{\bf $\ql$ bidder:} This is a traditional profit-maximizing bidder with no constraint. 
The dominant strategy for such a Bidder $j$ is to bid her value $v_{j,i}$ for impression $i$, regardless of how everyone else bids for that impression.

\medskip

{\bf $\tcpa$ bidder:} Such Bidder $j$ maximizes the number of conversions (i.e., the total value of the impressions which the bidder gets) subject to the constraint that the average cost per conversion is no greater than their $\tcpa$ $T_j\ge 0$. Namely, bidder $j$ solves the following maximization problem:
\begin{align}
    \max_{\forall i\in I,\, b_{j,i}\ge0,\,x_{j,i}\in[0,1]} \quad &  \sum_{i\in I \textrm{ s.t. } b_{j,i}\ge c_{j,i}(\b_{i},\r)} v_{j,i}x_{j,i}\nonumber\\
    \mbox{s.t. }\quad & \sum_{i\in I} c_{j,i}(\b_{i},\r) x_{j,i} \leq T_j \cdot \sum_{i\in I} v_{j,i} x_{j,i}\quad\forall j\in J\nonumber\\
    & x_{j,i} = 1 \mbox{ if } b_{j,i} > c_{j,i}(\b_{i},\r)\quad\forall j\in J,i\in I\label{con:x}.
\end{align}
\medskip

{\bf $\budgeted$ bidder:} Such Bidder $j$ maximizes the number of conversions subject to a budget constraint $B_j$. Namely, Bidder $j$ solves the following maximization problem:
\begin{align}
    \max_{\forall i\in I,\, b_{j,i}\ge0,\,x_{j,i}\in[0,1]} \quad &  \sum_{i\in I \textrm{ s.t. } b_{j,i}\ge c_{j,i}(\b_{i},\r)} v_{j,i}x_{j,i}\nonumber\\
    \mbox{s.t. }\quad & \sum_{i\in I} c_{j,i}(\b_{i},\r) x_{j,i} \leq B_j\quad\forall j\in J\nonumber\\
    & x_{j,i} = 1 \mbox{ if } b_{j,i} > c_{j,i}(\b_{i},\r)\quad\forall j\in J,i\in I\label{con:x1}.
\end{align}
\medskip

Notice that both $\tcpa$ bidder and $\budgeted$ bidder are allowed to decide the fraction of an impression they get in case they are tied for that impression (we say that bidder $j$ is {\bf tied} for an impression $i$ if $b_{j,i}=c_{j,i}(\b_i,\r)$). This is in line with the standard approach in the literature (e.g., budget pacing equilibrium~\cite{conitzer2021multiplicative}) that endogenizes the tie-breaking rule as part of the equilibrium concept which we will define shortly. Moreover, in the following proposition, we show that given other bidders' bids, it is optimal for a $\tcpa$ bidder (or a $\budgeted$ bidder) $j$ to bid uniformly\footnote{Qualitatively, this is same as the well-known result of~\citet{AggarwalBM19}. They prove this by introducing small perturbations to bidders' values. Instead, we take the approach that endogenizes the tie-breaking rule as part of the equilibrium concept, which is the standard approach in the literature for proving existence and computational complexity of equilibrium.}, i.e., the bids are characterized by a single {\bf bidding parameter} $\alpha_j\ge 0$ as follows: $\forall i\in I,\,b_{j,i}=\alpha_jv_{j,i}$.

\begin{proposition}\label{prop:uniform_bidding_optimal}
For a $\tcpa$ bidder (or a $\budgeted$ bidder resp.) $j$, the optimal bids for Problem~\eqref{con:x} (or~\eqref{con:x1} resp.) have the following form: there exists $\alpha_j\ge 0$ such that $\forall i\in I,\,b_{j,i}=\alpha_jv_{j,i}$.
\end{proposition}
The proof of Proposition~\ref{prop:uniform_bidding_optimal} is provided in appendix.

\subsection{Bidders' Subgame}

Bidders observe the reserve prices $\r = (r_k)_{k\in K}$ posted by the channels and decide their bids $\b_i(\r)$ for each impression $i$, and if a Bidder $j$ is tied for impression $i$, they can also decide the fraction $x_{j,i}(\r)$ of impression $i$ they get. In the previous subsection, we have shown that for any bidder of any type, the best response given other bidders' bids is bidding uniformly, and hence, we assume that each bidder $j$ uses uniform bidding with a bidding parameter $\alpha_j(\r)$.

Moreover, we assume that in the bidders' subgame, bidders use the {\em undominated uniform} bidding strategies. Specifically, for a $\ql$ bidder, bidding less than their value is dominated by bidding their value, and for a $\tcpa$ bidder $j$, using a bidding parameter less than $T_j$ is dominated by using a bidding parameter $\alpha_j(\r)=T_j$. To see the latter, notice that a $\tcpa$ bidder $j$ using a bidding parameter strictly less than $T_j$ cannot be $\tcpa$-constrained since their cost for any impression they are winning cannot be more than their bid $b_{j,i} = \alpha_j(\r) v_{j,i} < T_j v_{j,i}$. Thus, their $\tcpa$ constraint, i.e., the first constraint in Problem~\eqref{con:x}, is not tight, and hence, by increasing their bidding parameter to $T_j$, the bidder can only increase the total value without violating its $\tcpa$ constraint. In summary, we make the following assumption:

\begin{assumption}[Uniform Undominated Bidding]\label{lem:tcpa-bid-atleast-tcpa}
Each bidder $j\in J$ uses uniform bidding, i.e., $\forall i\in I,\,b_{j,i}(\r) = \alpha_j(\r) v_{j,i}$ for some bidding parameter $\alpha_j(\r)\ge 0$. Moreover, each $\ql$ bidder $j$ uses a bidding parameter $\alpha_j(\r) = 1$, and each $\tcpa$ bidder $j$ uses a bidding parameter $\alpha_j(\r)\geq T_j$.
\end{assumption}

The equilibrium solution we adopt for the bidders' subgame is a version of subgame perfection that takes into account endogenous tie-breaking rules, which is in line with the literature, e.g., the pacing equilibrium for $\budgeted$ bidders~\citep{conitzer2021multiplicative} and the autobidding equilibrium for $\tcpa$ bidders~\citep{li2022auto}.

\begin{definition}[Subgame Bidding Equilibrium]\label{def:uniform_bidding_equilibrium}
Consider the bidders' subgame given reserve prices $\r$ posted by the channels. An equilibrium for the subgame consists of bidders' bidding parameters $\boldsymbol{\alpha}(\r) = (\alpha_j(\r))_{j\in J}$ and probabilities of allocations of the impressions $\x(\r) = (x_{j,i}(\r))_{j \in J,i\in I}$ such that
\begin{enumerate}
    \item[(1)] Only a bidder whose bid is no less than the cost gets the impression: for $i\in I_k$, $x_{j,i}(\r)>0$ holds only if $b_{j,i}(\r)\ge c_{j,i}(\b_i(\r),\r)$.
    \item[(2)] Full allocation of any item with a bid above the reserve price: for $i\in I_k$, $\sum_{j\in J} x_{j,i}(\r)=1$ must hold if there exists some $\ell\in J$ such that $b_{\ell,i}(\r)> r_{\ell,i}$.
    \item[(3)] Constraints are satisfied: for each $j\in J_{\tiny \budgeted}$, $\sum_{i\in I}c_{j,i}(\b_i(r),\r)\cdot x_{j,i}(\r)\le B_j$, and for each $j\in J_{\tiny \tcpa}$, ${\sum_{i\in I}c_{j,i}(\b_i(\r),\r) x_{j,i}(\r)} \le T_j \cdot {\sum_{i\in I}v_{j,i}x_{j,i}(\r)}$.
    \item[(4)] For every Budgeted or tCPA bidder $j$, even if they can decide the fraction $x_{j,i}(\r)$ of an impression $i$ they get in case they are tied for impression $i$, increasing their bidding parameter would not increase their value without violating their budget/tCPA constraint.
\end{enumerate}
\end{definition}

The existence of subgame bidding equilibrium is a straightforward consequence by adapting the existence proofs of the pacing equilibrium for $\budgeted$ bidders~\citep{conitzer2021multiplicative} and the autobidding equilibrium for $\tcpa$ bidders~\citep{li2022auto}.
\begin{proposition}\label{prop:subgame_equilibrium_exists}
In the bidders' subgame, the subgame bidding equilibrium always exists.
\end{proposition}

\subsection{Channels}

We focus on two models that depend on the objective functions the channels may have: the {\em Local} channels model and the {\em Global} channels model.

{\bf Local Channels Model}: In this case, each channel sets its reserve price $r_k$ to maximize its own revenue given the other channels' reserve prices $\r_{-k}$. Thus, channel $k$ solves
$$ \max_{r_k} \sum_{j\in J}\sum_{i \in I_k} c_{j,i}(\b_i(r_k, \r_{-k}),r_k,\r_{-k})x_{j,i}(r_k,\r_{-k}).$$
\medskip

{\bf Global Channels Model}: In this case, the channels determine the reserve prices $\r$ to maximize the sum of the revenue across all channels. Thus, they set reserve prices solving
$$ \max_{\r} \sum_{k\in K} \sum_{j\in J}\sum_{i \in I_k} c_{j,i}(\b_i(\r),\r)x_{j,i}(\r).$$


\subsection{The Full Game}
We summarize the full game for the channels and the bidders as the following two-stage game:
\begin{itemize}
    \item[(S0)] Each Channel $k\in K$ chooses a uniform reserve price (in the cost-per-unit-value space) $r_k$ with finite precision\footnote{Note that assuming the reserve prices have finite precision is very natural for practice.} for their impressions $I_k$.
    \item[(S1)] Each Bidder $j\in J$ observes the reserve prices $\r$ posted by the channels (and the reserve prices $(p_i)_{i\in I}$ prechosen by the publishers), and then they choose a bidding parameter $\alpha_j(\r)$ and submit their bids according to $\alpha_j(\r)$ (see Assumption~\ref{lem:tcpa-bid-atleast-tcpa}). If Bidder $j$ is tied for impression $i$, they can also decide the fraction $x_{j,i}(\r)$ of impression $i$ they get.
\end{itemize}

By Proposition~\ref{prop:subgame_equilibrium_exists}, given any fixed $\r$ in the support of the channels' mixed strategies, stage (S1) has a subgame equilibrium between the bidders. We assume that stage (S1) always results into one such equilibrium deterministically, i.e., henceforth, we assume that $((\b_i(\r))_{i\in I},\x(\r))$ in stage (S1) is always a fixed subgame equilibrium given $\r$ (as defined in Definition~\ref{def:uniform_bidding_equilibrium}).

Channels are allowed to use mixed strategies in stage (S0), i.e., sampling their reserve price $r_k$ from a distribution $\mathcal{R}_k$. Notice that the game for the channels is a finite game between finite players, and hence, there always exists a mixed-strategy equilibrium by the celebrated Nash's theorem~\cite{nas1951non}.

Additionally, we assume the game is complete-information. That is,  $(v_{j,i}, B_j, T_j, p_i)_{j\in J, i\in I}$ are known to the channels and the bidders.

\subsection{Important Concepts}
We now present the main concepts which we will use to compare the outcomes of the local channels model to the global channels model.

\begin{definition}[Liquid Welfare]\label{def:liquid_wel}
The liquid welfare of a fractional allocation $\x =(x_{j,i})_{j\in J, i \in I}$ is
$$ {Wel}(\x) =  \sum_{j\in J_{\tiny \budgeted}}\min\left\{ \budget_j, \sum_{i\in I} v_{j,i} x_{j,i} \right\}+ \sum_{j \in J_{\tiny \tcpa }} \sum_{i\in I} T_jv_{j,i} x_{j,i} + \sum_{j \in J_{\tiny \ql }} \sum_{i\in I} v_{j,i} x_{j,i},$$
and the optimal liquid welfare is
$Wel^*:= \max_{\x \textrm{ that satisfies bidders' constraints}} Wel(\x)$.
\end{definition}

This concept of liquid welfare has been previously studied in e.g., \citet{AggarwalBM19} and \citet{azar2017liquid}, and it was first introduced by \citet{dobzinski2014efficiency}. It is well-known that optimal liquid welfare is an upper bound on the sum of the revenues of all channels. More precisely, optimal liquid welfare $Wel^*$ is greater or equal than the sum of the channels' revenues, which we denote by $Rev(\r):=\sum_{k\in K} \sum_{j\in J}\sum_{i \in I_k} c_{j,i}(\b_i(\r),\r)x_{j,i}(\r)$, regardless of the reserve prices they choose:

\begin{fact}
\label{fact:rev-wel}
For any $\r\in\R_{\ge0}^K$, $Rev(\r) \leq Wel^*$.
\end{fact}

Thus, we use the optimal liquid welfare as the benchmark to measure performance of the revenue in the local and global models. We let $\mbox{LocalEQ}$ denote the set that contains every mixed-strategy equilibrium $\mathcal{R}:=(\mathcal{R}_k)_{k\in K}$ for the channels in the local channel model, and we define the revenue guarantees in the local and global models as follows:

\begin{definition}[Revenue Guarantee]\label{def:rev_guarantee}
The revenue guarantees for the local and global models are defined as 
\begin{align*} 
RevG(Local) &= \frac{\inf_{\mathcal{R}\in \mbox{LocalEQ}} \E_{\r\sim\mathcal{R}}[\mbox{Rev}(\r)]}{Wel^*},\\
 RevG(Global) &= \frac{\sup_{\r\in\R_{\ge0}^K} \mbox{ Rev}(\r)}{Wel^*}.
\end{align*}
\end{definition}

Note that any reserve prices $\r$ in the support of any $\mathcal{R}\in\mbox{LocalEQ}$ in the local model is also feasible in the global model, thereby giving us the following fact.
\begin{fact}
\label{fact:local-global}
$RevG(Local) \leq RevG(Global) \leq 1$
\end{fact}

\medskip

Furthermore, to compare the outcomes of the two models, we use the standard notion of the Price of Anarchy~\cite{koutsoupias-poa}. 
\begin{definition}[Price of Anarchy]\label{def:poa}
The Price of Anarchy (PoA) of the local model compared to the global model is
$$\mbox{PoA} \;=\; \frac{\inf_{\mathcal{R}\in \mbox{LocalEQ}} \E_{\r\sim\mathcal{R}}[\mbox{Rev}(\r)]}{ \sup_{\r\in\R_{\ge0}^K} \mbox{ Rev}(\r)}.$$
\end{definition}
\section{Hardness of equilibrium computation}
\label{sec:hardness}
In this section, we study the computational complexity of computing equilibrium of our game. Our main result in this section is that we show that even just finding the subgame equilibrium (Definition~\ref{def:uniform_bidding_equilibrium}) for the bidders' subgame is already computationally hard:
\begin{theorem}
\label{thm:subgame-hardness}
Finding the subgame equilibrium (Definition~\ref{def:uniform_bidding_equilibrium}) is $\PPAD$-hard.
\end{theorem}

We prove this by reduction from the problem of finding an approximate Nash equilibrium for the 0-1 bimatrix game, which was shown to be $\PPAD$-hard in~\cite{CTV07}. The  basic  idea  of the proof is similar to that of the hardness result for finding a pacing equilibrium for budget-constrained quasi-linear bidders~\cite{CKK21}. However, we have to handle many difficulties that are unique to $\tcpa$ bidders.  Most notably, in contrast to budget-constrained quasi-linear bidders, whose bidding parameters are at most $1$, $\tcpa$ bidders do not have a natural upper bound for their bids, and their bidding parameters can be arbitrarily high when their $\tcpa$ constraints are not tight. We construct new gadgets that force $\tcpa$ bidders' bidding parameters to stay bounded but still leave a controlled amount of "slack" for them, so that they can bid on impressions that are more expensive than their $\tcpa$ but not win all of them. We provide the full reduction in the appendix.

Despite the computational hardness, we are able to prove tight revenue guarantees that channels can achieve in the equilibrium, which we will present in the subsequent sections.
\section{Revenue and Price of Anarchy with no Publisher Reserves}\label{zero-publisher-reserve}
In this section, we focus on the setting where impressions do not have publisher-chosen reserve prices, i.e. $p_i =0$ for every $i\in I$. We study the revenue guarantees that the channels can achieve in the local model where each channel chooses their reserve price out of their own self-interest vs. the global model where the channels cooperatively choose the reserve prices to maximize their total revenue. 

Our main results in this section are the following:
\begin{itemize}
    \item We establish a revenue guarantee (defined in Definition~\ref{def:rev_guarantee}) in the local model (Theorem~\ref{thm:rev-local}).
    \item Moreover, we prove that our revenue guarantee in the local model is \emph{tight even for the global model} (Proposition~\ref{ex:revg-ub}).
    \item Furthermore, as a corollary of the revenue guarantee in the local model, we immediately get a lower bound for the price of anarchy (Theorem~\ref{lem:ineq-POA}). We give a matching upper bound for the price of anarchy (Theorem~\ref{lem:tight-PoA}) and thus establish a \emph{tight separation between the local and global models}.
\end{itemize}

\subsection{Revenue Guarantees}

We begin by proving the main technical result of this section, which establishes a revenue guarantee for the local model. 
It is $\PPAD$-hard to actually compute the equilibrium, as shown in Theorem~\ref{thm:subgame-hardness}. Nevertheless, we will show that each channel can set a certain reserve price in order to guarantee itself a decent amount of revenue, irrespective of the reserve prices set by other channels.

To do this, we will show that each channel can set a reserve price which ensures that its revenue is at least a certain fraction of the total budget of
unconstrained $\budgeted$ bidders (Lemma~\ref{lem:budgeted-uniform} and Corollary~\ref{cor:min-rev-budg}). Then, we will show that each channel can set a reserve price which ensures that its revenue is at least a certain fraction of its contribution to the optimal liquid welfare (Definition~\ref{def:liquid_wel}) from $\tcpa$ and $\ql$ bidders (Lemma~\ref{lem:tcpa-ql} and Corollary~\ref{cor:tcpa-ql-rev}).
Finally, we will put these together to get the final revenue guarantee (Theorem~\ref{thm:rev-local}).

The main difficulty in this proof comes from $\budgeted$ bidders who are unconstrained, i.e. not spending their budget, at the equilibrium of the local model. The contribution to optimal Liquid Welfare from $\tcpa$ and $\ql$ bidders can be easily attributed to different channels (see the definition of $W_{\tiny \tcpa}^*(k)$ and $W_{\tiny \ql}^*(k)$ in Lemma~\ref{lem:tcpa-ql}) and there is a natural way for a channel to obtain a good fraction of its contribution as revenue (see Lemma~\ref{lem:tcpa-ql}). However, there is no obvious attribution for the contribution of $\budgeted$ bidders to different channels, and no obvious lower bound on the bid of $\budgeted$ bidders. In order to get a handle on unconstrained $\budgeted$ bidders, we define the notion of {\em Budget-fraction}. 
\begin{definition}[Budget-fraction $\beta_j$ and $\beta_{max}$, $\beta_{min}$]\label{def:budget-fraction}
For a $\budgeted$ bidder $j$, define their budget-fraction as $\beta_j = \frac{B_j}{\sum_{i\in I} v_{j,i}}$, i.e., the ratio of their budget to the sum of their values of all impressions.
Also, define $\beta_{max} = \max_{j \in J_{\tiny \budgeted}} \beta_j$ and $\beta_{min} = \min_{j \in J_{\tiny \budgeted}} \beta_j$.
\end{definition}
Intuitively, the budget-fraction for a $\budgeted$ bidder plays a role similar to the $\tcpa$ of a $\tcpa$ bidder. With this, we are ready to prove some key technical claims that will help establish a lower bound on the bids of unconstrained $\budgeted$ bidders, which in turn will help us find a good reserve price for these bidders.

\subsection*{Key Claims} 
\begin{claim}
\label{lem:unconstrained-win-all}
In a subgame equilibrium (Definition~\ref{def:uniform_bidding_equilibrium}), if a $\budgeted$ bidder $j$ is unconstrained, i.e. not spending all its budget, then they must be winning all impressions $i$ with $v_{j,i} > 0$ and cannot be tied with another unconstrained $\budgeted$ bidder on those impressions.
\end{claim}
\begin{proof}
Suppose for contradiction that a $\budgeted$ bidder $j$ is unconstrained but does not fully win certain impression $i$ (i.e., $x_{j,i}(\r)<1$) such that $v_{j,i}>0$. Notice that Bidder $j$ can increase the bidding parameter $\alpha_j$ without increasing the total spend until Bidder $j$ is tied for (but does not fully win) some impression $i'$ with $v_{j,i'}>0$. Such tie must occur, because otherwise, as Bidder $j$ increases $\alpha_j$, at some point Bidder $j$ will be tied for the impression $i$. However, this contradicts item (4) of Definition~\ref{def:uniform_bidding_equilibrium}, because Bidder $j$ can strictly increase the utility by increasing $x_{j,i'}(\r)$ by a sufficiently small amount such that their budget constraint is not violated.
\end{proof}
The next claim follows directly from Claim~\ref{lem:unconstrained-win-all}.
\begin{claim}
\label{cor:one-unbudgeted}
In a subgame equilibrium, for any impression $i\in I$, there can be at most one unconstrained $\budgeted$ bidder $j$ with $v_{j,i} > 0$.
\end{claim}

Next we prove the following claim, which will be helpful in bounding revenue against the optimal Liquid Welfare from unconstrained $\budgeted$ bidders. 
\begin{claim}
\label{lem:bid-atleast-budget-fraction}
If the final reserve price of an impression $i$ for a $\budgeted$ bidder $j$ satisfies that $r_{j,i}<\beta_j v_{j,i}$, then impression $i$ will be sold for a cost at least $r_{j,i}$ in the subgame equilibrium.
\end{claim}
\begin{proof}
We first show that unless impression $i$ is fully sold to bidder $j$ (in which case the statement holds trivially because the cost of impression $i$ is at least its reserve price $r_{j,i}$), Bidder $j$ will bid $b_{j,i} \geq \beta_j v_{j,i}$ for impression $i$.

Suppose $b_{j,i} < \beta_j v_{j,i}$ for contradiction. Recall that $\alpha_j$ denotes the bidding parameter of Bidder $j$. Since $b_{j,i} = \alpha_j v_{j,i}$ is assumed to be less than $\beta_j v_{j,i}$, we get $\alpha_j < \beta_j$.
Moreover, because the total amount spent by Bidder $j$ is at most the sum of their bids, we have that $$\mbox{Total amount spent by bidder } j \leq \sum_{i\in I} b_{j,i} = \sum_{i\in I} \alpha_j v_{j,i} < \sum_{i\in I} \beta_j v_{j,i} = B_j.$$
Thus, Bidder $j$ is unconstrained. By Claim~\ref{lem:unconstrained-win-all}, this bidder must be winning all its impressions.

Now we have shown that $b_{j,i}\ge\beta_j v_{j,i}$, which is by our assumption strictly greater than $r_{j,i}$. Thus, it follows from item (2) of Definition~\ref{def:uniform_bidding_equilibrium} that impression $i$ will be fully sold in the subgame equilibrium (for a cost that is at least $b_{j,i}>r_{j,i}$ because of item (1) of Definition~\ref{def:uniform_bidding_equilibrium}).
\end{proof}

The following claim, analogous to the claim above, will be used to bound revenue against the optimal Liquid Welfare from $\tcpa$ and $\ql$ bidders.
\begin{claim}
\label{lem:sold-at-rji}
If the final reserve price (i.e., in the cost space) of impression $i$ for a $\tcpa$ bidder $j$ (recall this is denoted by $r_{j,i}$ in the model section) satisfies that $r_{j,i}<T_j v_{j,i}$, then impression $i$ will be sold for a cost of at least $r_{j,i}$ in the subgame equilibrium. Similarly, if the final reserve price of impression $i$ for a $\ql$ bidder $j$ satisfies that $r_{j,i}<v_{j,i}$, then impression $i$ will be sold for a cost of at least $r_{j,i}$ in the subgame equilibrium.
\end{claim}
\begin{proof}
By Assumption~\ref{lem:tcpa-bid-atleast-tcpa}, a $\tcpa$ bidder $j$ bids $b_{j,i}(\r) \geq T_j v_{j,i}>r_{j,i}$ on impression $i$. Thus, by item (2) of Definition~\ref{def:uniform_bidding_equilibrium}, impression $i$ will be fully sold in the subgame equilibrium (for a cost that is at least $r_{j,i}$ because of item (1) of Definition~\ref{def:uniform_bidding_equilibrium}). An analogous argument holds for the case of $\ql$ bidder.
\end{proof}

Next, we first lower bound each channel's revenue against the optimal Liquid Welfare contribution from $\budgeted$ bidders (Lemma~\ref{lem:budgeted-uniform} and Corollary~\ref{cor:min-rev-budg}), and then we lower bound each channel's revenue against the welfare contribution from $\tcpa$ and $\ql$ bidders (Lemma~\ref{lem:tcpa-ql} and Corollary~\ref{cor:tcpa-ql-rev}). Finally, we will put these together to get a lower bound on the revenue guarantee (Theorem~\ref{thm:rev-local}). 

\subsection*{Welfare from $\budgeted$ Bidders}
\begin{lemma}
\label{lem:budgeted-uniform}
Let $E$ be any subgame equilibrium given any reserve prices. Define the following:
\begin{itemize}
    \item Let $J^E_C$ be the subset of $\budgeted$ bidders who are {\bf constrained}, i.e. are spending their entire budget in the equilibrium $E$.
    \item Let $J^E_U$ be the subset of $\budgeted$ bidders who are {\bf unconstrained}, i.e. are spending strictly less than their budget in the equilibrium $E$.
    \item For Channel $k$ and $\budgeted$ bidder $j$, let $\rho(k,j) = \frac{\sum_{i \in I_k} v_{j,i}}{\sum_{i\in I} v_{j,i}}$ be the ratio of the total value of impressions in $I_k$ for Bidder $j$ to the total value of all impressions in $I$ for Bidder $j$.
\end{itemize}
Then, for any $\varepsilon>0$,
\begin{enumerate}
    \item in the equilibrium $E$, the total revenue of all the channels from a Bidder $j \in J^E_C$ is no less than their budget $B_j$,
    \item and moreover, if Channel $k$ could set bidder-specific reserve prices $r_k(j) = (1-\varepsilon)\beta_j$ for each $\budgeted$ bidder $j$ (recall $\beta_j$ is the budget-fraction in Definition~\ref{def:budget-fraction}), then regardless of other channels' reserve prices, in the resulting subgame equilibrium (this is not necessarily $E$), Channel $k$ can obtain a revenue of at least $$\sum_{j \in J^E_U} (1-\varepsilon)\rho(k,j) B_j,$$
    \item and furthermore, Channel $k$ can set a uniform reserve price $r_k$ which is independent of $E$ such that regardless of other channels' reserve prices, in the resulting subgame equilibrium (not necessarily $E$), Channel $k$ will obtain a revenue of at least
    $$\frac{\sum_{j \in J^E_U} (1-\varepsilon)\rho(k,j) B_j} {2\max \big\{1,\big\lceil \log \frac{\beta_{max}}{\beta_{min}}\big\rceil\big\}}.$$
\end{enumerate}
\end{lemma}

\begin{proof}
\begin{enumerate}
\item Since bidders $j \in J^E_C$ are spending their entire budget in $E$ (by definition of $J^E_C$), the total revenue of all channels from them is equal to their budget.

\item Consider any equilibrium $E_r$ resulting from Channel $k$'s bidder-specific reserve prices given in the statement and arbitrary reserve prices of other channels (note $E_r$ is unrelated to $E$). Consider any impression $i \in I_k$. Since Channel $k$ has set a bidder-specific reserve price of $(1-\varepsilon)\beta_j$ for each $\budgeted$ bidder $j$, the reserve price of impression $i$ for $\budgeted$ bidder $j$ is $r_{j,i} =(1-\varepsilon)\beta_j v_{j,i}<\beta_j v_{j,i}$. Then, by Claim~\ref{lem:bid-atleast-budget-fraction}, each impression $i$ is sold for a price of at least $r_{j,i} = (1-\varepsilon)\beta_j v_{j,i}$ in the equilibrium $E_r$ for any $j\in J_{\tiny \budgeted}$. That is,
\begin{equation}\label{eq:revk_budget_frac}
    \textrm{the revenue of Channel $k$ in the equilibrium $E_r$} \geq \sum_{i \in I_k} \max_{j \in J_{\tiny \budgeted}} (1-\varepsilon)\beta_j v_{j,i}.
\end{equation}

Now let $I_k(j) \subseteq I_k$ denote the set of the impressions $i\in I_k$ such that $v_{j,i} > 0$. By Claim~\ref{cor:one-unbudgeted}, if $j$ and $j'$ are two bidders unconstrained in the equilibrium $E$, then $I_k(j)$ and $I_k(j')$ are disjoint.

Hence, we have that
\begin{align}
    \sum_{i \in I_k} \max_{j \in J_{\tiny \budgeted}} (1-\varepsilon)\beta_j v_{j,i}
    & \geq \sum_{j \in J^E_U} \sum_{i \in I_k(j)} (1-\varepsilon)\beta_j v_{j,i}  \label{eq:budg1_first}\\
    & = \sum_{j \in J^E_U} (1-\varepsilon)\beta_j \sum_{i \in I_k(j)} v_{j,i} \\
    & = \sum_{j \in J^E_U} (1-\varepsilon)\beta_j \rho(k,j) \sum_{i\in I} v_{j,i} \\
    & = \sum_{j \in J^E_U} (1-\varepsilon)\rho(k,j) B_j \label{eq:budg1_last},
\end{align}
which finishes the proof by Inequality~\eqref{eq:revk_budget_frac}.

\item The high-level idea for setting a good uniform reserve price is to bucketize the reserve prices and pick the one with the highest revenue potential. Specifically, we divide $\budgeted$ bidders into the following buckets:
$$J^s_{\tiny \budgeted} = \{j: j \in J_{\tiny \budgeted} \mbox{ and } 2^s \beta_{min} \leq \beta_j \leq 2^{s+1} \beta_{min}\}$$
for $s\in\{0\}\cup\left[\lceil \log \frac{\beta_{max}}{\beta_{min}} \rceil-1\right]$ (recall $\beta_{max},\beta_{min}$ are the largest and smallest budget-fractions respectively defined in Definition~\ref{def:budget-fraction}).
We observe that if Channel $k$ sets its uniform reserve price to $(1-\varepsilon)2^{s} \beta_{min}$, then for bidders $j \in J^s_{\tiny \budgeted}$, it holds that $r_{j,i}=(1-\varepsilon)2^{s} \beta_{min}v_{j,i}<\beta_j v_{j,i}$ for all impressions $i\in I_k$. Thus, by Claim~\ref{lem:bid-atleast-budget-fraction}, each impression $i\in I_k$ will get sold for a price of at least $\max_{j \in J^s_{\tiny \budgeted}} r_{j,i}=\max_{j \in J^s_{\tiny \budgeted}} (1-\varepsilon)2^{s} \beta_{min}v_{j,i} \geq \max_{j \in J^s_{\tiny \budgeted}} \frac{1-\varepsilon}{2} \beta_j v_{j,i}$ (the inequality is by bucketization) in the subgame equilibrium $E_r$ that results from Channel $k$ setting a uniform reserve price of $(1-\varepsilon)2^{s} \beta_{min}$ and arbitrary reserve prices set by other channels. Thus, the revenue of Channel $k$ from setting a uniform reserve price to $(1-\varepsilon)2^{s} \beta_{min}$
\begin{align}\label{eq:revk_per_bucket}
\mbox{Rev}_k((1-\varepsilon)2^{s} \beta_{min}) 
& \geq \sum_{i \in I_k} \max_{j \in J^s_{\tiny \budgeted}} \frac{1-\varepsilon}{2} \beta_j v_{j,i}.
\end{align}

Now let $I_k(j) \subseteq I_k$ denote the set of impressions $i\in I_k$ such that $v_{j,i} > 0$. Then, by Claim~\ref{cor:one-unbudgeted}, $I_k(j)$ and $I_k(j')$ are disjoint for two unconstrained $\budgeted$ bidders $j\neq j'$ in the equilibrium $E$. Hence, we have that
\begin{align}\label{eq:revk_sum_over_buckets}
    \sum_s \sum_{i \in I_k} \max_{j \in J^s_{\tiny \budgeted}} \frac{1-\varepsilon}{2} \beta_j v_{j,i}&\ge \sum_s \sum_{j \in J^E_U\cap J^s_{\tiny \budgeted}}\sum_{i \in I_k(j)} \frac{1-\varepsilon}{2} \beta_j v_{j,i}\nonumber\\
    &=\sum_{j \in J^E_U}\sum_{i \in I_k(j)} \frac{1-\varepsilon}{2} \beta_j v_{j,i}\nonumber\\
    &\ge\frac{1-\varepsilon}{2} \sum_{j \in J^E_U} \rho(k,j) B_j,
\end{align}
where the last inequality follows from the same derivation as in Inequalities~(\ref{eq:budg1_first}-\ref{eq:budg1_last}).

Finally, let $s^*=\argmax_{s\in\{0\}\cup\left[\lceil \log \frac{\beta_{max}}{\beta_{min}} \rceil-1\right]}\sum_{i\in I_k}\max_{j \in J^s_{\tiny \budgeted}} \frac{1-\varepsilon}{2} \beta_j v_{j,i}$. Then, we have that the revenue of Channel $k$ by setting a uniform reserve price $r_k^* = 2^{s^*} \beta_{min}$ (notice $r_k^*$ is indeed independent of $E$) is
\begin{align*}
\mbox{Rev}_k((1-\varepsilon)2^{s^*} \beta_{min}) 
& \geq \sum_{i \in I_k} \max_{j \in J^s_{\tiny \budgeted}} \frac{1-\varepsilon}{2} \beta_j v_{j,i} &&\text{(By Inequality~\eqref{eq:revk_per_bucket})}\\
&\ge \frac{\sum_s \sum_{i \in I_k} \max_{j \in J^s_{\tiny \budgeted}} \frac{1-\varepsilon}{2} \beta_j v_{j,i}}{\max\big \{1, \big \lceil \log \frac{\beta_{max}}{\beta_{min}}\big \rceil \big \}} &&\text{(By definition of $s^*$)}\\
&\ge \frac{\frac{1-\varepsilon}{2} \sum_{j \in J^E_U} \rho(k,j) B_j}{\max\big \{1, \big \lceil \log \frac{\beta_{max}}{\beta_{min}}\big \rceil \big \}} &&\text{(By Inequality~\eqref{eq:revk_sum_over_buckets})}.
\end{align*}
\end{enumerate}
\end{proof}

Item (3) in Lemma~\ref{lem:budgeted-uniform} implies the following corollary:
\begin{corollary}
\label{cor:min-rev-budg}
Define $\rho(k,j)$ as in Lemma~\ref{lem:budgeted-uniform}. Let $\mathcal{R}=(\mathcal{R}_k)_{k\in K}$ be any mixed-strategy equilibrium of the channels' game (i.e., S0), and let $E(\r)$ be the subgame equilibrium given any reserve prices $\r$ in the support of the channels' mixed strategies. Then, the expected revenue of Channel $k$ in the mixed-strategy equilibrium $\mathcal{R}$ is at least 
$$\mathbf{E}_{\r\sim \mathcal{R}}\left[\frac{\sum_{j \in J^{E(\r)}_U} (1-\varepsilon)\rho(k,j) \budget_j}{2\max\big \{1, \big \lceil \log \frac{\beta_{max}}{\beta_{min}}\big \rceil \big \}}\right].
$$
\end{corollary}

\subsection*{Welfare from $\tcpa$ and $\ql$ Bidders}
\begin{lemma}
\label{lem:tcpa-ql}
Let $\x^*$ be a welfare maximizing allocation (i.e., $\x^*$ is s.t. $Wel^*=Wel(\x^*)$ in Definition~\ref{def:liquid_wel}) 
and
\begin{itemize}
    \item let $W_{\tiny \tcpa}^*(k)$ be the liquid welfare generated by the impressions in $I_k$ allocated to $\tcpa$ bidders in $x^*$, i.e.,
    $$W_{\tiny \tcpa}^*(k):=\sum_{j \in J_{\tiny \tcpa }} \sum_{i\in I_k} T_jv_{j,i} x^*_{j,i},$$
    \item and let $W_{\tiny \ql}^*(k)$ be the liquid welfare generated by the impressions in $I_k$ allocated to quasi-linear bidders in $x^*$, i.e.,
    $$W_{\tiny \ql}^*(k):=\sum_{j \in J_{\tiny \ql }} \sum_{i\in I_k} v_{j,i} x^*_{j,i}.$$
\end{itemize}

Then, for any $\varepsilon>0$,
\begin{enumerate}
\item if Channel $k$ could set the bidder-specific reserve prices (also in the cost-per-unit-value space) $r_k(j)$ for each $\tcpa$ or $\ql$ bidder $j$ as follows:
\begin{equation*}
r_k(j) = \begin{cases}
(1-\varepsilon)T_j &\text{if $j$ is a $\tcpa$ bidder} \\
1-\varepsilon &\text{if $j$ is a $\ql$ bidder},
\end{cases}
\end{equation*}
then Channel $k$ obtains a revenue at least $(1-\varepsilon)(W_{\tiny \tcpa}^*(k) +W_{\tiny \ql}^*(k))$ regardless of what other channels do,

\item and moreover, we let $T_{max} = \max_{j \in J_{\tiny \tcpa}} T_j$ and let $T_{min} = \min_{j \in J_{\tiny \tcpa}} T_j$, and then Channel $k$ can set a uniform reserve price $r_k$ s.t. Channel $k$ obtains a revenue at least
$$\frac{(1-\varepsilon)(W_{\tiny \tcpa}^*(k) +W_{\tiny \ql}^*(k))}{ 2 + 2\max\big \{1, \big \lceil \log \frac{T_{max}}{T_{min}}\big \rceil \big \}}$$
regardless of what other channels do.
\end{enumerate}
\end{lemma}
\begin{proof}
\begin{enumerate}
    \item  Fix Channel $k$'s bidder-specific reserve prices as in the assumption and consider any subgame equilibrium. For any impression $i \in I_k$, let Bidder $j=\argmax_{\ell\in J_{\tiny \tcpa}\textrm{ s.t. } x^*_{\ell,i}>0} T_{\ell}v_{\ell,i}$, and let Bidder $q=\argmax_{\ell\in J_{\tiny \ql}\textrm{ s.t. } x^*_{\ell,i}>0} v_{\ell,i}$. Since $r_{j,i} =r_k(j)v_{j,i}= (1-\varepsilon)T_j v_{j,i}<T_j v_{j,i}$, it follows from Claim~\ref{lem:sold-at-rji} that impression $i$ will be sold for a cost of at least $r_{j,i}=(1-\varepsilon)T_j v_{j,i}$. Similarly, since $r_{q,i} =r_k(q)v_{q,i}= (1-\varepsilon)v_{q,i}<v_{q,i}$, it follows from Claim~\ref{lem:sold-at-rji} that impression $i$ will be sold for a cost of at least $r_{q,i}=(1-\varepsilon)v_{q,i}$. Moreover, note that the contribution of impression $i$ to $W_{\tiny \tcpa}^*(k)+W_{\tiny \ql}^*(k)$ is at most $\max\{v_{q,i},T_j v_{j,i}\}$, and we have shown impression $i$ will be sold for at least $(1-\varepsilon)$-fraction of this amount, it follows that channel $k$'s revenue is at least $(1-\varepsilon)(W_{\tiny \tcpa}^*(k)+W_{\tiny \ql}^*(k))$.
    
    \item The high-level idea for setting a good uniform reserve price is again to bucketize the bidder-specific reserve prices used above and set the uniform reserve price $r_k$ to the lower end of the bucket that has the highest revenue potential. Specifically, we divide all the $\tcpa$ bidders into the following buckets:
$$J^s_{\tiny \tcpa} = \{j: j \in J_{\tiny \tcpa}, 2^s T_{min} \leq T_j \leq 2^{s+1} T_{min}\}$$
for $s\in\{0\}\cup\left[\lceil \log \frac{T_{max}}{T_{min}} \rceil-1\right]$.

As before, for any impression $i \in I_k$, let bidder $j=\argmax_{\ell\in J_{\tiny \tcpa}\textrm{ s.t. } x^*_{\ell,i}>0} T_{\ell}v_{\ell,i}$ and bidder $q=\argmax_{\ell\in J_{\tiny \ql}\textrm{ s.t. } x^*_{\ell,i}>0} v_{\ell,i}$, and notice that the contribution of impression $i$ to $W_{\tiny \tcpa}^*(k)+W_{\tiny \ql}^*(k)$ is at most $\max\{v_{q,i},T_j v_{j,i}\}$.

If $T_j v_{j,i}>v_{q,i}$, let $s$ be such that $j \in J^s_{\tiny \tcpa}$. Suppose Channel $k$ sets a reserve price of $r_k = (1-\varepsilon)2^s T_{min}$ which is strictly less than $T_j$ because of the bucketization. Then, by Claim~\ref{lem:sold-at-rji}, impression $i$ will be sold at a cost at least $(1-\varepsilon)2^s T_{min} v_{j,i} \geq \frac{1-\varepsilon}{2} T_j v_{j,i}$ in the subgame equilibrium, where the inequality is because of the bucketization.

If $v_{q,i}\ge T_j v_{j,i}$, suppose Channel $k$ sets a reserve price $r_k = 1-\varepsilon$. Then, by Claim~\ref{lem:sold-at-rji}, impression $i$ will be sold at a cost at least $v_{q,i}$ in the subgame equilibrium.

Now we put these two cases together. Let $Rev_k(r_k)$ be the revenue of Channel $k$ at the subgame equilibrium if Channel $k$ sets a uniform reserve price $r_k$ (regardless of the reserve prices of other channels). Then, summing over all the buckets $s$, we have
$$\sum_{r_k\in\{1-\varepsilon,\,T_{min}\}\cup\left\{2^s T_{min}\mid\, s\in\left[\lceil \log \frac{T_{max}}{T_{min}} \rceil-1\right]\right\}} Rev_k(r_k) \geq \frac{1-\varepsilon}{2} (W_{\tiny \tcpa}^*(k)+W_{\tiny \ql}^*(k)),$$
because as we have shown in the above case analysis, all the buckets together cover at least $\frac{1-\varepsilon}{2}$-fraction of the liquid welfare of each impression's contribution to $W_{\tiny \tcpa}^*(k)+W_{\tiny \ql}^*(k)$.

Let $r^*_k = \argmax_{r_k\in\{1-\varepsilon,\,T_{min}\}\cup\left\{2^s T_{min}\mid\, s\in\left[\lceil \log \frac{T_{max}}{T_{min}} \rceil-1\right]\right\}} Rev_k(r_k)$. Then, by setting a reserve price of $r^*_k$, Channel $k$ can get a revenue of at least
$$\frac{(1-\varepsilon)(W_{\tiny \tcpa}^*(k) +W_{\tiny \ql}^*(k))}{ 2 + 2\max\big \{1, \big \lceil \log \frac{T_{max}}{T_{min}}\big \rceil \big \}}.$$
\end{enumerate}
\end{proof}

If Channel $k$ can always get certain amount of revenue by setting a particular uniform reserve price $r_k$ regardless of what other channels do, then Channel $k$'s revenue at any mixed-strategy equilibrium of the channels' game (i.e., stage (S0) of the full game) is at least the same amount (because otherwise Channel $k$ will deviate to the uniform reserve price $r_k$). Thus, item (2) in Lemma~\ref{lem:tcpa-ql} implies the following corollary:
\begin{corollary}
\label{cor:tcpa-ql-rev}
Let $W_{\tiny \tcpa}^*(k)$ and $W_{\tiny \ql}^*(k)$ be defined as in Lemma~\ref{lem:tcpa-ql} above. Then, for any $\varepsilon>0$, at any mixed-strategy equilibrium of the channels' game (S0), the expected revenue of channel $k$ is at least 
$$\frac{(1-\varepsilon)(W_{\tiny \tcpa}^*(k) +W_{\tiny \ql}^*(k))}{ 2 + 2\max\big \{1, \big \lceil \log \frac{T_{max}}{T_{min}}\big \rceil \big \}}.$$
\end{corollary}

\subsection*{The Final Revenue Guarantee}
\begin{theorem} 
\label{thm:rev-local}
For any $\varepsilon>0$,
\begin{equation*}
RevG(Local) \geq \frac{1-\varepsilon}{3 + 2\max\big \{1, \big \lceil \log \frac{T_{max}}{T_{min}}\big \rceil \big \} +2\max\big \{1, \big \lceil \log \frac{\beta_{max}}{\beta_{min}}\big \rceil \big \}}.
\end{equation*}

\end{theorem}
\begin{proof}
Let $\mathcal{R}=(\mathcal{R}_k)_{k\in K}$ be any mixed-strategy equilibrium of the channels' game, and let $E(\r)$ denote the subgame equilibrium given any reserve prices $\r$ in the support of the channels' mixed strategies. Let $\x^*$ be the liquid welfare maximizing allocation (i.e., $\x^*$ is s.t. $Wel^*=Wel(\x^*)$).

By Corollary~\ref{cor:tcpa-ql-rev}, the expected revenue of Channel $k$ in the equilibrium $\mathcal{R}$, denoted by $\mbox{Rev}_k[\mathcal{R}]$, is
$$\mbox{Rev}_k[\mathcal{R}]\ge \frac{(1-\varepsilon)(W_{\tiny \tcpa}^*(k) +W_{\tiny \ql}^*(k))}{ 2 + 2\max\big \{1, \big \lceil \log \frac{T_{max}}{T_{min}}\big \rceil \big \}},$$
where $W_{\tiny \tcpa}^*(k)$ and $W_{\tiny \ql}^*(k)$ are defined as in Lemma~\ref{lem:tcpa-ql}.

Thus, the expected total revenue of all channels, denoted by $\mbox{Rev}[\mathcal{R}]$, is
\begin{equation}
\label{eq:tcpa-ql}
\mbox{Rev}[\mathcal{R}]\ge\frac{(1-\varepsilon)(W_{\tiny \tcpa}^* + W_{\tiny \ql}^*)}{ 2 + 2\max\big \{1, \big \lceil \log \frac{T_{max}}{T_{min}}\big \rceil \big \}},
\end{equation}
where $W_{\tiny \tcpa}^*:=\sum_{k\in K}W_{\tiny \tcpa}^*(k)$ and $W_{\tiny \ql}^*:=\sum_{k\in K}W_{\tiny \ql}^*(k)$ denote the total contributions of $\tcpa$ and $\ql$ bidders to the liquid welfare of $x^*$ respectively.

Let $\rho(k,j)$ be as defined in Lemma~\ref{lem:budgeted-uniform}. Then, by Corollary~\ref{cor:min-rev-budg},
$$ \mbox{Rev}_k[\mathcal{R}] \geq \mathbf{E}_{\r\sim \mathcal{R}}\left[\frac{\sum_{j \in J^{E(\r)}_U} (1-\varepsilon)\rho(k,j) B_j}{2\max\big \{1, \big \lceil \log \frac{\beta_{max}}{\beta_{min}}\big \rceil \big \}}\right].
$$
Summing over all channels, we get 
\begin{equation}
\label{eq:nbc}    
\mbox{Rev}[\mathcal{R}] \geq \sum_k \mathbf{E}_{\r\sim \mathcal{R}}\left[\frac{\sum_{j \in J^{E(\r)}_U} (1-\varepsilon)\rho(k,j) B_j}{2\max\big \{1, \big \lceil \log \frac{\beta_{max}}{\beta_{min}}\big \rceil \big \}}\right] 
= \mathbf{E}_{\r\sim \mathcal{R}}\left[\frac{\sum_{j \in J^{E(\r)}_U} (1-\varepsilon)B_j}{2\max\big \{1, \big \lceil \log \frac{\beta_{max}}{\beta_{min}}\big \rceil \big \}}\right].
\end{equation}

Also, by item (1) of Lemma~\ref{lem:budgeted-uniform}, we have that
\begin{equation}
\label{eq:bc}
\mbox{Rev}[\mathcal{R}] \geq \mathbf{E}_{\r\sim \mathcal{R}}\left[\sum_{j \in J^{E(\r)}_C} B_j\right].
\end{equation}

Notice that 
$$Wel^* \leq W_{\tiny \tcpa}^* + W_{\tiny \ql}^* + \sum_{j \in J_{\tiny \budgeted}} B_j,$$
and then the theorem follows from Inequalities~\eqref{eq:tcpa-ql},~\eqref{eq:nbc} and~\eqref{eq:bc}.
\end{proof}

Combining Theorem~\ref{thm:rev-local} with Fact~\ref{fact:local-global}, we get the following corollary:
\begin{corollary}
\label{cor:rev-global}
For any $\varepsilon>0$, 
\begin{equation*}
RevG(Global) \geq \frac{1-\varepsilon}{3 + 2\max\big \{1, \big \lceil \log \frac{T_{max}}{T_{min}}\big \rceil \big \} +2\max\big \{1, \big \lceil \log \frac{\beta_{max}}{\beta_{min}}\big \rceil \big \}}.
\end{equation*}
\end{corollary}
\bigskip

Finally, we show that the above revenue guarantees in the local and global models are both tight up to a constant factor by constructing an example using the well-known "equal-revenue" trick.

\begin{proposition}\label{ex:revg-ub}
For the single-channel setting, there is an instance where $RevG(Global) = RevG(Local)=O(1/(\log(T_{max}/T_{min}) +\log( \beta_{max}/\beta_{min})))$.
\end{proposition}
\begin{proof}
Since there is only one channel, $RevG(Global) = RevG(Local)$.

Consider $2^{\ell}$ $\tcpa$ bidders with $\tcpa$s $1/2^{\ell}$ for $\ell=0, \ldots, w_1-1$, each interested in a unique impression with a value of $1$ (i.e. their value for every other impression is 0, and everyone else's value for their impression is 0). Similarly, there are $2^{\ell}$ $\budgeted$ bidders with budgets $1/2^{\ell}$ for $\ell=0, \ldots, w_2-1$, each interested in a unique impression with a value of $1$ (i.e. their value for every other impression is 0, and everyone else's value for their impression is 0). Optimal liquid welfare is $w_1 + w_2$ obtained by giving everyone their unique impression. The best uniform reserve price cannot get a revenue more than $4$. This shows that $RevG(Global) \leq 4 / (w_1+w_2) = 4 / (2 + \log \frac{T_{max}}{T_{min}} + \log \frac{\beta_{max}}{\beta_{min}})$.
\end{proof}

\subsection{Price of Anarchy}
In this subsection, we study how much total revenue the channels lose in the local model where they set their uniform reserve prices out of their own self-interest compared to the global model where they choose the reserve prices cooperatively. Specifically, we consider the standard notion -- price of anarchy $PoA$ (Definition~\ref{def:poa}).
First, we observe that the revenue guarantee from Theorem~\ref{thm:rev-local} immediately implies a lower bound for the $PoA$:
\begin{theorem}\label{lem:ineq-POA}
For any $\varepsilon>0$,
\begin{equation*}
     PoA \geq \frac{1-\varepsilon}{3 + 2\max\big \{1, \big \lceil \log \frac{T_{max}}{T_{min}}\big \rceil \big \} +2\max\big \{1, \big \lceil \log \frac{\beta_{max}}{\beta_{min}}\big \rceil \big \}}.
\end{equation*}
\end{theorem}
\begin{proof}
By definition of $PoA$ (Definition~\ref{def:poa}), $PoA=\frac{RevG(Local)}{RevG(Global)}$. By Fact~\ref{fact:local-global}, $RevG(Global)\le 1$. It follows that $PoA\ge RevG(Local)$, and then the proof finishes by applying Theorem~\ref{thm:rev-local}.
\end{proof}

Next, we show that the $PoA$ lower bound in Theorem~\ref{lem:ineq-POA} is tight (up to a constant factor).

\begin{theorem}\label{lem:tight-PoA}
There is an instance with two channels such that $PoA=O(1/(\log(T_{max}/T_{min}) +\log( \beta_{max}/\beta_{min})))$.
\end{theorem}
\paragraph{The high-level idea:} We first construct an ``equal-revenue'' instance (which consists of many $\tcpa$ bidders $J_1$ with geometrically decreasing $\tcpa$s, each interested in a unique impression owned by Channel $k_1$) as in the proof of Proposition~\ref{ex:revg-ub}. For this ``equal-revenue'' instance, Channel $k_1$ cannot simultaneously get good revenues from all the bidders in $J_1$ by setting a uniform reserve price.

Now the key idea is to introduce another Channel $k_2$ and another $\tcpa$ bidder $j_2\notin J_1$, such that Channel $k_2$ only owns one impression, for which only bidder $j_2$ has strictly positive value. Moreover, Bidder $j_2$ has a value for each impression $i$ in channel $k_1$, and Bidder $j_2$'s value for impression $i$ is carefully chosen to be proportional to the $\tcpa$ of the bidder in $J_1$ who is interested in impression $i$. Thus, if Bidder $j_2$ makes a uniform bid (in the cost-per-unit-value space), it results into non-uniform bids (in the cost space) for the impressions in Channel $k_1$, which are proportional to the $\tcpa$s of bidders in $J_1$. We can think of these non-uniform bids as non-uniform bidder-specific reserve prices for bidders in $J_1$, which are proportional to their $\tcpa$s. Thus, we are able to extract the full revenue from all the bidders in $J_1$ (similar to item (1) of Lemma~\ref{lem:tcpa-ql}).

Finally, we just need to argue the above idea can only be successfully applied in the global model but not in the local model. This is because in the local model, Channel $k_2$ sets a high reserve price for its sole impression in order to profit more from bidder $j_2$, and as a result, Bidder $j_2$ does not have enough ``slack'' to make a sufficiently high uniform bid to incur sufficiently high bidder-specific reserve prices for bidders in $J_1$.

The construction of the instance with $\budgeted$ bidders uses essentially the same idea as above. The full proof is provided in Appendix~\ref{app:sec5}.


As a corollary of Theorem~\ref{lem:ineq-POA} and Theorem~\ref{lem:tight-PoA}, we have the following tight price of anarchy:
\begin{theorem}[Price of Anarchy]
\label{thm:poa-main}
$PoA=\Theta(1/(\log(T_{max}/T_{min}) +\log( \beta_{max}/\beta_{min})))$.
\end{theorem}
\section{Price of Anarchy with Publisher Reserves}\label{publisher-reserve}

This section studies the general version of the model where a publisher, owning impression $i$, sets a minimum price $p_i$ for the impression to be sold.\footnote{Recall that the price $p_i$ is also in the cost-per-unit-value space.} The main finding we obtain is that Theorem~\ref{thm:poa-main} dramatically depends on not having publisher prices. We show that with publisher prices and general channels,  $PoA = 0$ in the worst case (Theorem~\ref{th:general-publisher-model}). 

We then restrict our attention to an important subclass of instances where channels are {\em scaled} copy of each other. That is, channels share a set of a homogeneous set of impressions and differ on the revenue share each owns. In this context, we show that $PoA$ has non-trivial lower bound only if there is one bidder in the auction. In this case, $PoA = 1/|K|$, and hence, depends on the number of channels in the game in contrast to our results in Section~\ref{zero-publisher-reserve}.




\subsection*{General Channels}

We now present the main result of the section for the general case when channels can have arbitrary asymmetries for the impressions they own with arbitrary publisher reserve prices.

\begin{theorem}\label{th:general-publisher-model}
If publishers can set arbitrary minimum prices on their impressions, then there is an instance for which $PoA = 0$.
\end{theorem}

\begin{proof}
Consider the following instance with two channels and one bidder who is a $\tcpa$ bidder with a target constraint $T=1$. Channel 1 has only one impression to sell. This impression does not have any publisher pricing constraint ($p_i=0$). Channel 2 has $q$ impressions to sell, each of these impressions has the same publisher pricing constraint $p_i = 1+ 1/q$. The bidder's valuation for all impressions is the same, i.e., $v_i = 1$ for all $i\in I$.

We assert that in the global model, it is optimal to set reserve prices equal to zero for both channels. Indeed, with no reserve prices, the bidder can purchase all impressions since she gets a value of $1+q$ for a total cost of $q\cdot (1+1/q) = 1+q$. This is the optimal solution for the global model as the total revenue is exactly the optimal liquid welfare.  

On the other hand, in the local model, it is a strictly dominant strategy for Channel 1 to set a uniform reserve $r_1=1$: if $r_1>1$, Channel 1 gets zero revenue. If $r_1\leq 1$, the bidder purchases its impression, which leads to a revenue of $r_1$. Thus, Channel 1 strictly prefers to set a reserve price of $r_1=1$. Because of $r_1=1$, the bidder cannot afford to buy any impression of Channel 2 since the cost of each impression is at least $1+1/q$. Thus, in this equilibrium, the bidder submits a uniform bid of $1$, gets only the impressions sold by Channel 1, and the global revenue is $1$.  

Therefore from this instance we have that $RevG(Local)/RevG(Global)\leq 1/(1+q)$. We conclude the proof by taking $q\to \infty$.
\end{proof}

The intuition behind the previous result comes from instances where some of the channels have high publisher prices relative to the bidder's $\tcpa$ targets while some other channels do not have publisher prices. In these instances, in the global model, channels benefit by keeping low reserve prices in the {\em cheap} channels (without publisher reserves) as they provide subsidy to the $\tcpa$ bidders to buy impressions from the {\em expensive} channels. However, when the cheap channels are myopic, they would like to raise their reserve prices to increase their local revenue. This local behavior negatively impacts the revenue of the expensive channels, which in turn, is negative for all channels.

Given that the reason for the previous negative $PoA$ result is the asymmetry of the publisher prices on the different channels, in what follows we restrict our $PoA$ analysis for a special subclass where channels are {\em scaled} versions of each other.

\subsection*{Scaled Channels}

The scaled channels model consists of weights $\gg=(\gamma_1,\ldots,\gamma_k) \in \Delta([0,1]^K)$ \footnote{$\Delta([0,1]^K$ is the unit simplex in $\R^K$} so that Channel $k$ owns a fraction $\gamma_k$ of each impression $i\in I$.\footnote{For simplicity of the exposition we assume that impressions are divisible. A similar model with non-divisible impressions would assume that each impression $i$ is duplicated so that Channel $k$ owns a fraction $\gamma_k$ of those duplicates.}


 The first result shows that, surprisingly, so long as there are more than one bidder participating in the auctions, then $PoA = 0$ in the worst case.
 
 \begin{theorem}\label{th: scaled channels PoA=0}
 For the scaled channels models if there are two or more bidders participating in the auctions, then there is an instance for which $PoA = 0$.
 \end{theorem}
 
The instance we construct (deferred to Appendix~\ref{ap:sec-publisher}) consists of two channels and two $\tcpa$ bidders. The idea of the instance is that the main source of revenue for the channels comes from Bidder 1 buying the expensive impressions, those with high publisher reserve price. Bidder 1 needs enough slack to be able to purchase those expensive impressions. Thus, Bidder 1 needs to buy enough cheap impressions. However, the cheap impressions may have a high price if Bidder 2 sets a high bid. Bidder 2 can only set a high bid if, instead, it has enough slack from (other) cheap impressions. The crux of the argument is that, in the global model, by setting a sufficiently high reserve price, the channels can avoid Bidder 2 to have enough slack. This, in turn, allows Bidder $1$ to have slack to buy the expensive impressions. On the contrary, for the local models, there is an equilibrium where both channels set a low reserve. This prevents Bidder 1 to buy expensive impressions because Bidder 2 is setting a high bid and removing Bidder 1's slack.

As a corollary of this instance, we show that in the autobidding framework setting a high reserve price like in the global model not only increases revenue but also increases the welfare. This contrasts with the classic profit-maximizing framework where there is a negative correlation between high reserve prices and welfare.



We finish this section by showing that for the case of only one bidder participating across all channels the $PoA$ is always strictly positive (for pure-strategy equilibria).

\begin{theorem}\label{poa: 1/k}
If there is only a single bidder, then for pure-strategy equilibria we have that $PoA =\frac {1}{ |K|}$, where $|K|$ is the number of channels in the game.
\end{theorem}

We defer the proof to Appendix~\ref{ap:sec-publisher}. We note that in contrast to the results of Section~\ref{zero-publisher-reserve} where the $PoA$ is independent of the number of channels, in the setting with publisher reserves, the $PoA$ directly depends of the number of channels.
\section{Further Discussion}\label{sec: further-discussions}
In this paper, we have established tight bounds on \emph{revenue} guarantees and Price of Anarchy when the reserve prices are set in the \emph{cost-per-unit-value} space. Two natural follow-up questions are:
\begin{quote}
    \begin{itemize}
        \item Can we obtain similar bounds for \emph{welfare} of the bidders?
        \item What are the revenue guarantees if the channels set reserve prices in the \emph{cost-per-impression} space?
    \end{itemize}
\end{quote}

We briefly discuss how to extend some of our results to answer these questions. We defer the details to the full paper.
\subsection*{Bounds for Welfare}
Most of our revenue and Price of Anarchy results carry over to {\em welfare}. In particular for the setting without publisher reserves, we can get bounds similar to the the revenue bounds in Theorem~\ref{thm:rev-local} and Proposition~\ref{ex:revg-ub} and the Price of Anarchy bound in Theorem~\ref{thm:poa-main} for welfare (see Appendix~\ref{app:welfare} for a proof sketch). Many of the results in the setting without publisher reserves also carry over to welfare. We defer the details to the full paper.

We also observe an interesting phenomena -- in contrast to the quasi-linear setting, using a higher reserve price can sometimes increase the welfare (see the discussion after Theorem~\ref{th: scaled channels PoA=0}).

\subsection*{Uniform cost-per-impression reserve prices}
 We can obtain a revenue guarantee analogous to Theorem~\ref{thm:rev-local} when channels set uniform cost-per-impression reserve prices (i.e., value-independent and the same for all bidders and impressions). We can do this by adapting the bucketization arguments in Section~\ref{zero-publisher-reserve} to bucketize $T_j v_{j,i}$ instead of $T_j$ for $\tcpa$ bidders, bucketize $v_{j,i}$ for Quasi-linear bidders, and bucketize $\beta_j v_{j,i}$ instead of $\beta_j$ for $\budgeted$ bidders.

\bibliographystyle{ACM-Reference-Format}
\bibliography{bibliography}

\appendix

\section{Bidding Uniformly is Optimal}

\begin{proof}[Proof of Proposition~\ref{prop:uniform_bidding_optimal}]
We prove the proposition using a greedy-exchange argument. We will maintain the invariant $\forall i\in I,\,b_{j,i}=\alpha_jv_{j,i}$. Initially, we let $\alpha_j=0$ and $\forall i\in I,\,x_{j,i}=0$, and then we update them using a greedy procedure:
Until the first constraint in Problem~\eqref{con:x} (or~\eqref{con:x1} resp.) is tight for solution $(b_{j,i})_{i\in I}, (x_{j,i})_{i\in I}$ or $\forall i\in I \textrm{ s.t. } v_{j,i}>0, x_{j,i}=1$ (i.e., bidder $j$ has already won every impression for which they have strictly positive value), do the following
\begin{enumerate}
    \item[1.] if bidder $j$ is tied for an impression $i$ and $x_{j,i}<1$ (when there are multiple such impressions, choose an arbitrary one), increase $x_{j,i}$ until $x_{j,i}=1$ or the stop condition is met,
    \item[2.] and if there is no such impression, increase $\alpha_j$ until bidder $j$ is tied for a new impression $i$ and go to Step 1.
\end{enumerate}

Notice that whenever the above procedure increases $x_{j,i_1}$ for an impression $i_1$ in Step 1, two things must hold: (i) for any impression $i_2\in I$ such that $\frac{c_{j,i_2}(\b_{i_2},\r)}{v_{j,i_2}}<\alpha_j$, we have $x_{j,i_2}=1$ (because for any such $i_2$, in the the above procedure, bidder $j$ would have been tied for $i_2$ before $i_1$, and the above procedure would have already increased $x_{j,i_2}$ to $1$), and (ii) $\frac{c_{j,i_1}(\b_{i_1},\r)}{v_{j,i_1}}=\alpha_j$ (because the procedure only increases $x_{j,i_1}$ when bidder $j$ is tied for impression $i_1$). Therefore, at any moment when the above procedure is increasing $x_{j,i}$ for some impression $i$, impression $i$ must be the current ``best bang for the buck'', i.e., among all the impressions $\ell\in I$ such that $x_{j,\ell}<1$, impression $i$ has the smallest cost-per-unit-value $\frac{c_{j,i}(\b_{i},\r)}{v_{j,i}}$ for bidder $j$.

Now we let $(b^*_{j,i})_{i\in I}, (x^*_{j,i})_{i\in I}$ be the solution to Problem~\eqref{con:x} (or~\eqref{con:x1} resp.) that the above greedy procedure converges to and let $(b'_{j,i})_{i\in I}, (x'_{j,i})_{i\in I}$ be any feasible solution to Problem~\eqref{con:x} (or~\eqref{con:x1} resp.), and we want to show that $(b^*_{j,i})_{i\in I}, (x^*_{j,i})_{i\in I}$ is not worse than $(b'_{j,i})_{i\in I}, (x'_{j,i})_{i\in I}$.

To this end, we rank all the impressions in $I$ according to their cost-per-unit-value $\frac{c_{j,i}(\b_{i},\r)}{v_{j,i}}$ for bidder $j$ in the increasing order $\pi$, i.e., $\pi$ is a permutation over $I$ such that $\frac{c_{j,\pi(i_1)}(\b_{\pi(i_1)},\r)}{v_{j,\pi(i_1)}}\le\frac{c_{j,\pi(i_2)}(\b_{\pi(i_2)},\r)}{v_{j,\pi(i_2)}}$ for any $i_1<i_2$. Let $i_0$ be the smallest number such that $x'_{j,\pi(i_0)}\neq x^*_{j,\pi(i_0)}$. It must hold that $x^*_{j,\pi(i_0)}> x'_{j,\pi(i_0)}$. To see this, notice that if $x^*_{j,\pi(i_0)}<1$, because the greedy procedure prioritize increasing $x^*_{j,\pi(i_0)}$ over any other $x^*_{j,\pi(i)}$ for $i>i_0$ until the first constraint in Problem~\eqref{con:x} (or~\eqref{con:x1} resp.) is tight, and $x'_{j,\pi(i)}=x^*_{j,\pi(i)}$ for all $i<i_0$ by definition of $i_0$, we must have $x^*_{j,\pi(i_0)}\ge x'_{j,\pi(i_0)}$ (otherwise $(x'_{j,i})_{i\in I}$ should violate the first constraint). If $x^*_{j,\pi(i_0)}=1$, then $x^*_{j,\pi(i_0)}\ge x'_{j,\pi(i_0)}$ holds trivially. Since in both cases we have $x^*_{j,\pi(i_0)}\ge x'_{j,\pi(i_0)}$, and we assumed that $x'_{j,\pi(i_0)}\neq x^*_{j,\pi(i_0)}$, it follows that $x^*_{j,\pi(i_0)}> x'_{j,\pi(i_0)}$.

Let $i'>i_0$ be such that $x'_{j,\pi(i')}>0$. There must exist such $i'$ WLOG, because otherwise it is obvious that $(b^*_{j,i})_{i\in I}, (x^*_{j,i})_{i\in I}$ is the better solution. Now consider the overall cost-per-unit-value $T'=\frac{\sum_{i\in I} c_{j,i}(\b_{i},\r) x'_{j,i}}{\sum_{i\in I} v_{j,i} x'_{j,i}}$ and the total cost $B'=\sum_{i\in I} c_{j,i}(\b_{i},\r) x'_{j,i}$. Because $\frac{c_{j,\pi(i_0)}(\b_{\pi(i_0)},\r)}{v_{j,\pi(i_0)}}\le\frac{c_{j,\pi(i')}(\b_{\pi(i')},\r)}{v_{j,\pi(i')}}$, if we decrease $x'_{j,i'}$ by $\frac{\delta}{v_{j,\pi(i')}}$ and increase $x'_{j,i_0}$ by $\frac{\delta}{v_{j,\pi(i_0)}}$ for $\delta=\min\{v_{j,\pi(i_0)}(x^*_{j,i_0}-x'_{j,i_0}),\, v_{j,\pi(i')}x'_{j,i'}\}$, then neither $T'$ nor $B'$ can increase, and the total value $\sum_{i\in I} v_{j,i} x'_{j,i}$ does not change. (Note that such change for $x'_{j,i'}$ and $x'_{j,i_0}$ is feasible after changing the bids  $b'_{j,i'}$ and $b'_{j,i_0}$ appropriately.)

Furthermore, we can repeat the above argument whenever there exists $i_0\in I$ such that $x'_{j,\pi(i_0)}\neq x^*_{j,\pi(i_0)}$ to make $x'_{j,\pi(i_0)}= x^*_{j,\pi(i_0)}$, which shows that $(x^*_{j,i})_{i\in I}$ achieves better (or equal) total value than $(x'_{j,i})_{i\in I}$.
\end{proof}

\section{Proof of Hardness of Finding Subgame Equilibrium}\label{app:proof-of-hardness}
In this section, we prove that it is $\PPAD$-hard to find the subgame equilibrium (Definition~\ref{def:uniform_bidding_equilibrium}) even when the subgame only consists of $\tcpa$ bidders and does not have reserve prices. Since we only consider $\tcpa$ bidders and no reserve prices in this section, we first simplify the notion of the subgame equilibrium by restricting to $\tcpa$ bidders in the following subsection.

\subsection{Subgame Equilibrium for $\tcpa$ Bidders}
Without reserve prices, the subgame equilibrium for $\tcpa$ bidders can be simplified as the following \emph{uniform-bidding equilibrium}, which is essentially same as the autobidding equilibrium in~\cite{li2022auto}, and hence, we also refer to the subgame for $\tcpa$ bidders as \emph{uniform-bidding game} in this section.
\begin{definition}[Uniform-Bidding Equilibrium for $\tcpa$ Bidders]\label{def:pacing_equilibrium}
In the subgame with $m$ items (impressions) $I$ and $n$ $\tcpa$ bidders with $\tcpa$s $T_1,\dots,T_n$, let $\alpha=(\alpha_1,\dots,\alpha_n)$ be the vector of the bidders' bidding parameters, where $\alpha_j\ge T_j$, and let $x=(x_{1,1},\dots,x_{n,m})$ be the vector of the allocations of the items, where $x_{j,i}\in [0,1]$ is the fraction of item $i$ being allocated to bidder $j$, and thus $\sum_{j\in[n]} x_{ji}\le 1$, and let $p_i$ be the second highest bid for item $i$, and thus bidder $j$ pays $p_ix_{j,i}$ for item $i$. We say $(\alpha, x)$ is a uniform-bidding equilibrium if
\begin{enumerate}
    \item[(1)] Only the bidder with highest bid gets the item: $x_{j,i}>0$ only if $\alpha_jv_{j,i}\ge \alpha_{\ell}v_{\ell,i}$ for all $\ell\in[n]$.
    \item[(2)] Full allocation of any item with a positive bid: $\sum_{j\in[n]} x_{j,i}=1$ if $\alpha_{\ell}v_{\ell,i}>0$ for some $\ell\in[n]$.
    \item[(3)] $\tcpa$s are satisfied: for each $j\in[n]$, $\frac{\sum_{i\in[m]}p_i x_{j,i}}{\sum_{i\in[m]}v_{j,i}x_{j,i}}\le T_j$.
    \item[(4)] Every bidder's bidding parameter is such that even if they can decide the fraction of an item they get in case of a tie, increasing their bidding parameter would not increase their total value without violating their $\tcpa$ constraint.
\end{enumerate}
\end{definition}

\subsection{Hardness of Finding the Uniform-Bidding Equilibrium}\label{app:hardness}
We reduce computing an (approximate) mixed-strategy Nash equilibrium of a 0-1 (win-lose) bimatrix game to computing a uniform-bidding equilibrium of the uniform-bidding game for $\tcpa$ bidders. The basic idea of the reduction is similar to that of the hardness result for finding uniform-bidding equilibrium for budget-constrained quasi-linear bidders~\cite{CKK21}. However, we do have to handle many difficulties that are unique to the $\tcpa$ constraints. Most notably, in contrast to budget-constrained quasi-linear bidders, whose bidding parameter is at most $1$, $\tcpa$ bidders do not have a natural upper bound for their bids, and their bidding parameters can be arbitrarily high when their $\tcpa$ constraints are not binding.

We start by defining the 0-1 bimatrix game and the approximate Nash equilibrium of this game.
\begin{definition}[0-1 bimatrix game $\cG(A,B)$]
In a 0-1 bimatrix game, there are two players, and they both have $n$ strategies to choose from. Player 1's cost matrix is $A\in\{0,1\}^{n\times n}$, i.e., player 1's cost is $A_{ij}$ if player 1 plays the $i$-th strategy, and player 2 plays the $j$-th strategy. Similarly, player 2's cost matrix is $B\in\{0,1\}^{n\times n}$, i.e., player 2's cost is $B_{ij}$ if player 1 plays the $i$-th strategy, and player 2 plays the $j$-th strategy..
\end{definition}
\begin{definition}[$\eps$-approximate Nash equilibrium]
In a 0-1 bimatrix game $\cG(A,B)$, suppose that player 1 plays mixed strategy $x\in[0,1]^n$ s.t. $1^Tx=1$, and player 2 plays mixed strategy $y\in[0,1]^n$ s.t. $1^Ty=1$. Then, we say $(x,y)$ is an $\eps$-approximate Nash equilibrium if it holds for all $z\in[0,1]^n$ s.t. $1^Tz=1$ that
\begin{align*}
    x^TAy&\le z^TAy+\eps,\\
    x^TBy&\le x^TAz+\eps.
\end{align*}
\end{definition}

Finding an approximate Nash equilibrium for 0-1 bimatrix game was shown to be $\PPAD$-hard~\cite{CTV07}.

\begin{lemma}[{\citet[Theorem~6.1]{CTV07}}]\label{lem:bimatrix}
For any constant $c>0$, finding $1/n^c$-approximate Nash equilibrium for 0-1 bimatrix game is $\PPAD$-hard.
\end{lemma}

Now, given a 0-1 bimatrix game $\cG(A,B)$ with arbitrary cost matrices $A$ and $B$, we construct a uniform-bidding game $\cI(A,B)$ for $\tcpa$ bidders as follows.

\subsubsection{Construction of Hard Instance $\cI(A,B)$}
\begin{description}
\item[Bidders:] For each player $p\in\{1,2\}$ and each strategy $s\in[n]$, we introduce two $\tcpa$ bidders $\cC(p,s)$ and $\cD(p,s)$. In addition, we have two more $\tcpa$ bidders $\cT_1$ and $\cT_2$.
\item[Items:] For each $p\in\{1,2\}$ and each $s\in[n]$, we construct an \emph{expensive item} $H(p,s)$, a \emph{cheap item} $L(p,s)$, a set of \emph{normalized items} $\{N(p,s)_i\mid i\in[n]\}$, and a set of \emph{expenditure items} $\{E(p,s)_i\mid i\in[n]\}$ for bidder $\cC(p,s)$, and moreover, we construct a cheap item $D(p,s)$ for bidder $\cD(p,s)$. Furthermore, we have a special item $T$ for bidders $\cT_1$ and $\cT_2$ and a cheap item $T_2$ for bidder $\cT_2$.
\item[Valuations:] We first give an informal description of the valuations, and then we provide the formal definition. Let $\delta_1=1/n^2$ and $\delta_2=1/n^4$.

Bidder $\cC(1,s)$ has high values (i.e., $n^3$) for the items $H(1,s)$ and $L(1,s)$, medium values (i.e., $3$) for their own normalized items $\{N(1,s)_i\mid i\in[n]\textrm{ and } i\neq s\}$, low values (i.e., $1$) for their own expenditure items $\{E(1,s)_i\mid i\in[n]\}$, normalized item $N(1,s)_s$, and bidder $\cC(1,t)$'s $s$-th normalized item $N(1,t)_s$, and negligible values (i.e., $\delta_1B_{st}$) for bidder $\cC(2,t)$'s $s$-th expenditure item $E(2,t)_s$. Bidder $\cC(2,s)$'s valuation is analogous.

On the other hand, bidder $\cD(1,s)$ has the same value (i.e., $1$) as bidder $\cC(1,s)$ for bidder $\cC(1,s)$'s $s$-th normalized item $N(1,s)_s$ and value $1$ for their own item $D(p,s)$.

Bidder $\cT_1$ has the same value (i.e., $n^3$) as bidder $\cC(1,s)$ for bidder $\cC(1,s)$'s expensive item $H(1,s)$ and value $1$ for the item $T$. Bidder $\cT_2$ has the same value (i.e., $1$) for the item $T$ as bidder $\cT_1$ and value $1$ for their own item $T_2$.

Formally, we use the notation $v(\textrm{bidder},\textrm{item})$ to denote a bidder's value of an item.
For all $p\in\{1,2\}$, all distinct $s,t\in[n]$ and all $i\in[n]$, we let
\begin{align*}
v(\cC(p,s),N(p,s)_s)=1&,\,\,v(\cC(p,s),N(p,s)_t)=3\\
v(\cC(p,s),N(p,t)_s)=1&,\,\,v(\cC(p,s),E(p,s)_i)=1,\\
v(\cC(1,s),E(2,i)_s)=\delta_1B_{si}&,\,\,v(\cC(2,t),E(1,i)_t)=\delta_1A_{it},\\
v(\cC(p,s),H(p,s))=n^3&,\,\, v(\cC(p,s),L(p,s))=n^3,\\
v(\cD(p,s),D(p,s))=1&,\,\, v(\cD(p,s),N(p,s)_s)=1.
\end{align*}
Moreover, we let $v(\cT_1,H(p,s))=n^3$ for all $p\in\{1,2\}$ and $s\in[n]$, and $v(\cT_1,T)=v(\cT_2,T)=v(\cT_2,T_2)=1$. For any other $(\textrm{bidder},\textrm{item})$ pair that did not appear above, $v(\textrm{bidder},\textrm{item})=0$.
\item[$\tcpa$s:] Bidder $\cT_1$'s $\tcpa$ is $1$, and bidder $\cT_2$'s $\tcpa$ is $\delta_2$. For all $p\in\{1,2\}$ and $s\in[n]$, bidder $\cD(p,s)$'s $\tcpa$ is $\delta_2$, and

\begin{align*}
&\textrm{bidder $\cC(1,s)$'s $\tcpa$}=
\frac{n^3+n+\delta_1\sum_{t\in[n]}A_{st}+3/2}{2n^3+4n-2},\\
&\textrm{bidder $\cC(2,s)$'s $\tcpa$}=
\frac{n^3+n+\delta_1\sum_{t\in[n]}B_{ts}+3/2}{2n^3+4n-2}.
\end{align*}
\end{description}

\subsubsection{Proof of Hardness}
\begin{theorem}\label{thm:hardness}
It is $\PPAD$-hard to find a uniform-bidding equilibrium in uniform-bidding game for $\tcpa$ bidders.
\end{theorem}
We prove Theorem~\ref{thm:hardness} by showing that if we find a uniform-bidding equilibrium for our hard instance $\cI(A,B)$, then we also find a $O(1/n)$-approximate Nash equilibrium for the 0-1 bimatrix game $\cG(A,B)$ (the theorem follows by Lemma~\ref{lem:bimatrix}). We split the proof of the theorem into a series of lemmata.

\begin{lemma}\label{lem:C_pacing_param_ge_1}
In any uniform-bidding equilibrium of $\cI(A,B)$, for all $p\in\{1,2\}$ and $s\in[n]$, bidder $\cC(p,s)$'s bidding parameter is at least $1$.
\end{lemma}
\begin{proof}
Suppose for contradiction bidder $\cC(p,s)$'s bidding parameter is strictly less than $1$. Then, bidder $\cC(p,s)$ does not get any fraction of the item $H(p,s)$, because bidder $\cT_1$ has the same value for the item $H(p,s)$ as bidder $\cC(p,s)$, and bidder $\cT_1$'s bidding parameter is at least bidder $\cT_1$'s $\tcpa$, which is $1$. Now let us upper bound the CPA (\emph{cost-per-acquisition}, i.e., bidder's total payment divided by bidder's total value) of bidder $\cC(p,s)$. 

First, the total value that bidder $\cC(p,s)$ gets is at least the value of the item $L(p,s)$, which is $n^3$, because no one other than $\cC(p,s)$ has positive value for $L(p,s)$, and hence $\cC(p,s)$ always wins $L(p,s)$ for free.

Moreover, the total payment that bidder $\cC(p,s)$ makes is at most $\cC(p,s)$'s bidding parameter times $\cC(p,s)$'s total value of the items except $H(p,s)$ and $L(p,s)$, because $\cC(p,s)$ does not get any fraction of $H(p,s)$ and gets $L(p,s)$ for free. It is straightforward to verify that by construction of $\cI(A,B)$, $\cC(p,s)$'s total value of the items except $H(p,s)$ and $L(p,s)$ is less than $5n$. Since we assume for contradiction that $\cC(p,s)$'s bidding parameter is less than $1$, the total payment $\cC(p,s)$ makes is less than $5n$.

Thus, bidder's $\cC(p,s)$'s CPA is less than $5n/n^3$, which is much less than $\cC(p,s)$'s $\tcpa$. Next, we show that this contradicts the fourth property in Definition~\ref{def:pacing_equilibrium}. Specifically, we can first assume WLOG that bidder $\cC(p,s)$ does not tie for any item, because otherwise $\cC(p,s)$ can increase the bidding parameter by an arbitrarily small amount such that $\cC(p,s)$ gets the full item which $\cC(p,s)$ ties for, and $5n$ would still be an upper bound of $\cC(p,s)$'s total payment (by the same argument as before), which contradicts the fourth property in Definition~\ref{def:pacing_equilibrium}. Then, we notice there exist items for which bidder $\cC(p,s)$ has positive value such as $H(p,s)$. Therefore, if $\cC(p,s)$ raises the bidding parameter until the first time $\cC(p,s)$ ties for a new item for which $\cC(p,s)$ has positive value, then because $\cC(p,s)$ already gets positive value with a CPA that is much less than $\cC(p,s)$'s $\tcpa$, $\cC(p,s)$ can afford at least a fraction of that new item (which contradicts the fourth property in Definition~\ref{def:pacing_equilibrium}).
\end{proof}

\begin{lemma}\label{lem:matching_pacing_param}
In any uniform-bidding equilibrium of $\cI(A,B)$, bidder $\cT_2$'s bidding parameter is equal to bidder $\cT_1$'s bidding parameter, and bidder $\cD(p,s)$'s bidding parameter is equal to bidder $\cC(p,s)$'s bidding parameter for all $p\in\{1,2\}$ and $s\in[n]$.
\end{lemma}
\begin{proof}
First, we show that in a uniform-bidding equilibrium, bidder $\cT_2$'s bidding parameter is equal to bidder $\cT_1$'s bidding parameter. Notice that no one other than bidder $\cT_2$ has positive value for the item $T_2$, and thus, $\cT_2$ gets $T_2$ with value $1$ for free, which gives $\cT_2$ the flexibility to afford certain fraction of the item $T$ regardless of its price, because bidder $\cT_2$'s $\tcpa$ is positive. Since bidder $\cT_2$ has the same value (i.e., $1$) for the item $T$ as bidder $\cT_1$ and can always afford a fraction of $T$, it follows by the fourth property in Definition~\ref{def:pacing_equilibrium} that in a uniform-bidding equilibrium, $\cT_2$'s bidding parameter should be no less than $\cT_1$'s bidding parameter. On the other hand, if $\cT_2$'s bidding parameter is strictly greater than $\cT_1$'s bidding parameter, which is at least $1$, then $\cT_2$ will win the full item $T$ for a price that is at least $1$, and it follows that $\cT_2$'s CPA is at least $1/2$, which is much higher than $\cT_2$'s $\tcpa$ (i.e., $\delta_2$). Thus, $\cT_2$'s bidding parameter is no greater than (and hence equal to) $\cT_1$'s bidding parameter.

The proof of the equivalence between bidder $\cD(p,s)$'s bidding parameter and bidder $\cC(p,s)$'s is similar. Specifically, $\cD(p,s)$ also has a free item $D(p,s)$ with value $1$, and $\cD(p,s)$ also has a positive but tiny $\tcpa$ (i.e., $\delta_2$), and thus, $\cD(p,s)$ can afford certain fraction of the item $N(p,s)_s$. Notice that $\cD(p,s)$ and $\cC(p,s)$ have the same value (i.e., $1$) for the item $N(p,s)_s$, and $\cC(p,s)$'s bidding parameter is no less than $\cC(p,s)$'s $\tcpa$ ($\approx1/2$). The rest of the proof is same as the proof above for bidder $\cT_2$ and bidder $\cT_1$.
\end{proof}

\begin{lemma}\label{lem:T1_pacing_param_is_1}
In any uniform-bidding equilibrium of $\cI(A,B)$, bidder $\cT_1$'s bidding parameter is $1$.
\end{lemma}
\begin{proof}
$\cT_1$'s bidding parameter is at least $1$, because $\cT_1$'s $\tcpa$ is $1$. It suffices to prove that $\cT_1$'s bidding parameter is at most $1$.

Now suppose for contradiction, bidder $\cT_1$'s bidding parameter is strictly greater than $1$. In the proof of Lemma~\ref{lem:matching_pacing_param}, we have shown that bidder $\cT_2$'s bidding parameter is equal to $\cT_1$'s bidding parameter and that $\cT_2$ can not afford the full item $T$. Therefore, $\cT_1$ must get a fraction of $T$ by the second property of Definition~\ref{def:pacing_equilibrium}, and the payment per value $\cT_1$ makes for $T$ is exactly $\cT_1$'s bidding parameter, which is strictly greater than $1$ by our assumption for contradiction. Moreover, for all $p\in\{1,2\}$ and $s\in[n]$, (i) by Lemma~\ref{lem:C_pacing_param_ge_1}, bidder $\cC(p,s)$'s bidding parameter is at least $1$, and (ii) bidder $\cT_1$ has the same value for the item $H(p,s)$ as bidder $\cC(p,s)$ by construction of $\cI(A,B)$. Hence, if bidder $\cT_1$ wins any fraction of the item $H(p,s)$ for any $p\in\{1,2\}$ and $s\in[n]$, the payment per value $\cT_1$ makes for $H(p,s)$ is at least $1$. Therefore, overall, bidder $\cT_1$'s CPA is strictly greater than $1$ and thus violates $\cT_1$'s $\tcpa$, which is a contradiction.
\end{proof}

\begin{lemma}\label{lem:D_gets_le_delta_2}
In any uniform-bidding equilibrium, for all $p\in\{1,2\}$ and $s\in[n]$, bidder $\cC(p,s)$ wins at least $1-2\delta_2$ fraction of the item $N(p,s)_s$.
\end{lemma}
\begin{proof}
Bidder $\cC(p,s)$ and bidder $\cD(p,s)$ tie for the item $N(p,s)_s$, because they have the same value $1$ for this item and the same bidding parameter by Lemma~\ref{lem:matching_pacing_param}. Thus, the payment per value for the item $N(p,s)_s$ is equal to $\cC(p,s)$'s bidding parameter which is $\ge1$. Since the only other item bidder $\cD(p,s)$ gets is $D(p,s)$ (value $1$ and zero cost), and $\cD(p,s)$ has $\tcpa$ $\delta_2$, it follows by straightforward calculation that $\cD(p,s)$ can afford no more than $2\delta_2$ fraction of the item $N(p,s)_s$. By the second property in Definition~\ref{def:pacing_equilibrium}, $\cC(p,s)$ wins at least $1-2\delta_2$ fraction of $N(p,s)_s$.
\end{proof}

\begin{lemma}\label{lem:C_pacing_param_le_3}
In any uniform-bidding equilibrium of $\cI(A,B)$, for all $p\in\{1,2\}$ and $s\in[n]$, bidder $\cC(p,s)$'s bidding parameter is strictly less than $3$.
\end{lemma}
\begin{proof}
We prove the lemma for bidder $\cC(1,s)$ (the case of bidder $\cC(2,s)$ is analogous).
Suppose for contradiction bidder $\cC(1,s)$'s bidding parameter is at least $3$, we show that $\cC(1,s)$'s $\tcpa$ must be violated. To this end, we count the total value $\cC(1,s)$ gets and the total payment $\cC(1,s)$ makes.

First, bidder $\cC(1,s)$ is the only bidder who has postive value for the item $L(1,s)$. Thus, bidder $\cC(1,s)$ gets value $n^3$ from the item $L(1,s)$ with zero payment.

Notice that bidder $\cT_1$ and bidder $\cC(1,s)$ have the same value $n^3$ for the item $H(1,s)$, and by Lemma~\ref{lem:T1_pacing_param_is_1}, bidder $\cT_1$'s bidding parameter is $1$ which is strictly less than bidder $\cC(1,s)$'s bidding parameter ($\ge3$), and hence, bidder $\cC(1,s)$ wins the full item $H(1,s)$ with payment $n^3$ and gets value $n^3$.

Bidder $\cC(1,s)$ and bidder $\cD(1,s)$ have the same value $1$ for the item $N(1,s)_s$. Lemma~\ref{lem:matching_pacing_param} shows that bidder $\cD(1,s)$'s bidding parameter is equal to bidder $\cC(1,s)$'s bidding parameter ($\ge3$). Therefore, the price per value of the item $N(1,s)_s$ is at least $3$. By Lemma~\ref{lem:D_gets_le_delta_2}, $\cC(1,s)$ gets at least $1-2\delta_2$ fraction of the item $N(1,s)_s$ and hence pays at least $3(1-2\delta_2)$ for that fraction of $N(1,s)_s$. That is, $\cC(1,s)$ gets at most value $1$ from $N(1,s)_s$ (because $\cC(1,s)$'s value for the full item $N(1,s)_s$ is $1$) and pays at least $3(1-2\delta_2)$.

The other items for which bidder $\cC(1,s)$ has positive value are $\{E(1,s)_t\mid t\in[n]\}$, $\{N(1,s)_t\mid t\in[n]\textrm{ and } t\neq s\}$, 
$\{E(2,t)_s\mid t\in[n]\}$ and $\{N(1,t)_s\mid t\in[n]\textrm{ and } t\neq s\}$.

Because bidder $\cC(2,t)$ has value $\delta_1A_{st}$ for the item $E(1,s)_t$, and by Lemma~\ref{lem:C_pacing_param_ge_1} $\cC(2,t)$'s bidding parameter is at least $1$, if bidder $\cC(1,s)$ wins the full item $E(1,s)_t$, the payment $\cC(1,s)$ makes is at least $\delta_1A_{st}$. For $t\neq s$ and $t\in[n]$, because bidder $\cC(1,t)$ has value $1$ for the item $N(1,s)_t$, and $\cC(1,t)$'s bidding parameter is at least $1$ by Lemma~\ref{lem:C_pacing_param_ge_1}, if bidder $\cC(1,s)$ wins the full item $N(1,s)_t$, the payment $\cC(1,s)$ makes is at least $1$. We can assume WLOG that $\cC(1,s)$ gets the full value of $E(1,s)_t$ (i.e., $1$) with payment $\delta_1A_{st}$ for each $t\in[n]$ and gets the full value of $N(1,s)_t$ (i.e., $3$) with payment $1$ for each $t\neq s$, because these are the best payments per value $\cC(1,s)$ can hope for these items, and these payments per value are much lower than $\cC(1,s)$'s $\tcpa$ ($\approx1/2$). Namely, if $\cC(1,s)$'s CPA does not exceed $\cC(1,s)$'s $\tcpa$, giving the items $\{E(1,s)_t\mid t\in[n]\}$ and $\{N(1,s)_t\mid t\in[n]\textrm{ and } t\neq s\}$ to $\cC(1,s)$ and charging the above payments per value will not violate $\cC(1,s)$'s $\tcpa$. Therefore, WLOG bidder $\cC(1,s)$ gets value $4n-3$ from the items $\{E(1,s)_t\mid t\in[n]\}$ and $\{N(1,s)_t\mid t\in[n]\textrm{ and } t\neq s\}$ and pays $\sum_{t\in[n]}\delta_1A_{st}+n-1$.

On the other hand, because bidder $\cC(2,t)$ has value $1$ and bidder $\cC(1,s)$ has value $\delta_1B_{st}$ for the item $E(1,t)_s$, and by Lemma~\ref{lem:C_pacing_param_ge_1} $\cC(2,t)$'s bidding parameter is at least $1$, if bidder $\cC(1,s)$ wins any fraction of the item $E(1,t)_s$, the payment per value $\cC(1,s)$ makes is at least $1/(\delta_1B_{st})$ (or bidder $\cC(1,s)$ never wins any fraction of this item if $B_{st}=0$). For $t\neq s$ and $t\in[n]$, because bidder $\cC(1,t)$ has value $3$ and bidder $\cC(1,s)$ has value $1$ for the item $N(1,t)_s$, and $\cC(1,t)$'s bidding parameter is at least $1$ by Lemma~\ref{lem:C_pacing_param_ge_1}, if bidder $\cC(1,s)$ wins any fraction of the item $N(1,t)_s$, the payment per value $\cC(1,s)$ makes is at least $3$. Notice that the payments per value for these items are all much higher than $\cC(1,s)$'s $\tcpa$, and hence, we can assume WLOG $\cC(1,s)$ does not win any fraction of these items. Namely, if bidder $\cC(1,s)$'s CPA does not exceed $\cC(1,s)$'s $\tcpa$, taking the items $\{E(2,t)_s\mid t\in[n]\}$ and $\{N(1,t)_s\mid t\in[n]\textrm{ and } t\neq s\}$ away from bidder $\cC(1,s)$ will not violate $\cC(1,s)$'s $\tcpa$. Therefore, WLOG bidder $\cC(1,s)$ gets zero value from the items $\{E(2,t)_s\mid t\in[n]\}$ and $\{N(1,t)_s\mid t\in[n]\textrm{ and } t\neq s\}$ and pays zero.

At the end of each paragraph above, we stated the values bidder $\cC(1,s)$ gets from different items and the associated payments. In summary, bidder $\cC(1,s)$'s total value is at most $2n^3+4n-2$, and the total payment is at least $n^3+3(1-2\delta_2)+\sum_{t\in[n]}\delta_1A_{st}+n-1$. Thus, $\cC(1,s)$'s CPA exceeds $\cC(1,s)$'s $\tcpa$, which is a contradiction.
\end{proof}

\begin{lemma}\label{lem:exists_C_pacing_param_exceed_1}
In any uniform-bidding equilibrium, for all $p\in\{1,2\}$, there exists $s\in[n]$ such that bidder $\cC(p,s)$'s bidding parameter is strictly greater than $1+1/n$.
\end{lemma}
\begin{proof}
We prove the lemma for $p=1$ (the case of $p=2$ is analogous). Suppose for contradiction that for all $s\in[n]$, bidder $\cC(1,s)$'s bidding parameter is in $[1,1+1/n]$ (we know that it is at least $1$ by Lemma~\ref{lem:C_pacing_param_ge_1}), we upper bound bidder $\cC(1,s)$'s CPA.

First, bidder $\cC(1,s)$ gets the item $L(1,s)$ for free, since there is no competition for this item. Moreover, bidder $\cC(1,s)$ can get a fraction of the item $H(p,s)$, since $\cC(1,s)$ and $\cT_1$ have the same value $n^3$ for this item, and by Lemma~\ref{lem:T1_pacing_param_is_1} $\cT_1$'s bidding parameter is $1$ which is no larger than $\cC(1,s)$'s. Let $\tau\in[0,1]$ denote the fraction of the item $H(p,s)$ which bidder $\cC(1,s)$ wins, and hence $\cC(1,s)$ gets value $\tau n^3$ from $H(p,s)$ and pays $\tau n^3$. Furthermore, For each $t\in[n]$, bidder $\cC(1,s)$ gets the full item $E(1,s)_t$ and pays at most $3\delta_1A_{st}$, because bidder $\cC(2,t)$ has value $\delta_1A_{st}$ for this item, and $\cC(2,t)$'s bidding parameter is less than $3$ by Lemma~\ref{lem:C_pacing_param_le_3}. For each $t\in[n]$ and $t\neq s$, bidder $\cC(1,s)$ gets the full item $N(1,s)_t$ and pays at most $1+1/n$, because bidder $\cC(1,t)$ has value $1$ for this item, and $\cC(1,t)$'s bidding parameter is $\le1+1/n$ by our assumption. Finally, by Lemma~\ref{lem:D_gets_le_delta_2}, $\cC(1,s)$ wins at least $1-2\delta_2$ fraction of $N(1,s)_s$ and pays at most $1+1/n$ (because $\cC(1,s)$'s bidding parameter times $\cC(1,s)$'s value for the full item $N(1,s)_s$ is $\le1+1/n$). In addition, it is easy to verify that with bidding parameter $\le1+1/n$, bidder $\cC(1,s)$ can not get any fraction of the items $\{E(2,t)_s\mid t\in[n]\}$ and $\{N(1,t)_s\mid t\in[n]\textrm{ and } t\neq s\}$.

In summary, bidder $\cC(1,s)$ gets total value at least $(1+\tau)n^3+4n-2-2\delta_2$ and makes total payment at most $\tau n^3+(n-1)(1+1/n)+\sum_{t\in[n]}3\delta_1A_{st}+1$. Therefore, the resulting upper bound of bidder $\cC(1,s)$'s CPA is maximized when $\tau=1$, and the maximum is
$$\frac{n^3+n-1/n+\sum_{t\in[n]}3\delta_1A_{st}+1}{2n^3+4n-2-2\delta_2},$$
which is strictly less than $\cC(1,s)$'s $\tcpa$. Therefore, if $\cC(1,s)$ raises the bidding parameter until the first time $\cC(1,s)$ ties for a new item (such item exists, e.g., $\{N(1,t)_s\mid t\in[n]\textrm{ and } t\neq s\}$), $\cC(1,s)$ can afford a fraction of that item, which contradicts the fourth property in Definition~\ref{def:pacing_equilibrium}. (One might notice that raising $\cC(1,s)$'s bidding parameter will break the tie for the item $N(1,s)_s$ between $\cC(1,s)$ and $\cD(1,s)$, but this is not an issue, because in the above calculation for the upper bound of the total payment made by $\cC(1,s)$, we already take the payment for the full item $N(1,s)_s$ into account.)
\end{proof}

Now we let $\tilde{x}_s$ denote bidder $\cC(1,s)$'s bidding parameter and let $\tilde{y}_t$ denote a bidder $\cC(2,t)$'s bidding parameter in a uniform-bidding equilibrium. We define a probability vector $x=(x_1,\dots,x_n)$ which corresponds to a mixed strategy for player 1 and a probability vector $y=(y_1,\dots,y_n)$ which corresponds to a mixed strategy for player 2 in the bimatrix game $\cG(A,B)$ as follows
\begin{align*}
x_s = \frac{\tilde{x}_s-1}{\sum_{i\in[n]} (\tilde{x}_i-1)},\,\,y_t = \frac{\tilde{y}_t-1}{\sum_{i\in[n]} (\tilde{y}_i-1)}.
\end{align*}
Note that $x$ is a valid probability vector, because for all $i\in[n]$, $\tilde{x}_i\ge 1$ by Lemma~\ref{lem:C_pacing_param_ge_1}, and there exists $i\in[n]$ such that $\tilde{x}_i>1$ by Lemma~\ref{lem:exists_C_pacing_param_exceed_1}. Similarly, $y$ is also a valid probability vector. The next lemma implies that $(x,y)$ is an $O(1/n)$-approximate Nash equilibrium of $\cG(A,B)$, which proves Theorem~\ref{thm:hardness}, because $(x,y)$ obviously can be computed efficiently from the uniform-bidding equilibrium of $\cI(A,B)$, and finding an $O(1/n)$-approximate Nash equilibrium of $\cG(A,B)$ is $\PPAD$-hard in general by Lemma~\ref{lem:bimatrix}.

\begin{lemma}\label{lem:best_strategy_exceed_1}
For all $s\in[n]$, if bidder $\cC(1,s)$'s bidding parameter $\tilde{x}_s$ is strictly greater than $1$, then the $s$-th strategy is a $O(1/n)$-approximate best response\footnote{We say a pure strategy is an $\eps$-approximate best response to the other player's mixed strategy if the expected cost of this pure strategy is at most the expected cost of any other pure strategy plus $\eps$.} for player $1$ to player 2's mixed strategy $y$ in the 0-1 bimatrix game $\cG(A,B)$. Similarly, if bidder $\cC(2,t)$'s bidding parameter $\tilde{y}_t$ is strictly greater than $1$, then the $t$-th strategy is a $O(1/n)$-approximate best response for player $2$ to player 1's mixed strategy $x$ in $\cG(A,B)$.
\end{lemma}
\begin{proof}
We prove the first part of the lemma, i.e., if $\tilde{x}_s>1$, then the $s$-th strategy is an $O(1/n)$-approximate best response to $y$. (The other part is analogous.) Formally, we want to show for all $s'\in[n]$ and $s'\neq s$, $\sum_{t\in[n]}A_{st}y_t\le \sum_{t\in[n]}A_{s't}y_t+O(1/n)$. By definition of $y_t$, this inequality is equivalent to
\begin{align*}
\sum_{t\in[n]}A_{st}(\tilde{y}_t-1)&\le \sum_{t\in[n]}A_{s't}(\tilde{y}_t-1)+O(1/n)\sum_{i\in[n]} (\tilde{y}_i-1).
\end{align*}
By Lemma~\ref{lem:C_pacing_param_ge_1} and Lemma~\ref{lem:exists_C_pacing_param_exceed_1}, $\sum_{i\in[n]} (\tilde{y}_i-1)$ is at least $1/n$. Hence, it suffices to prove that for all $s'\in[n]$ and $s'\neq s$,
\begin{align}\label{eq:approximate_best_response}
\sum_{t\in[n]}A_{st}(\tilde{y}_t-1)&\le \sum_{t\in[n]}A_{s't}(\tilde{y}_t-1)+O(1/n^2).
\end{align}
First of all, we show that for all $i\in[n]$, bidder $\cC(1,i)$'s $\tcpa$ must be binding WLOG. Specifically, by Lemma~\ref{lem:C_pacing_param_le_3}, bidder $\cC(1,i)$'s bidding parameter is less than $3$, which implies that $\cC(1,i)$ does not get any fraction of the items $\{N(1,t)_i\mid t\in[n]\textrm{ and } t\neq i\}$. However, if $\cC(1,i)$'s $\tcpa$ is not binding, $\cC(1,i)$ can raise the bidding parameter and afford certain fraction of those items, which contradicts the fourth property in Definition~\ref{def:pacing_equilibrium}. The only exception is that $\cC(1,i)$ might have only won a fraction of the item $N(1,i)_i$ because of a tie with bidder $\cD(1,i)$ (or the item $H(p,s)$ in case of a tie with bidder $\cT_1$), and then raising the bidding parameter might violate $\cC(1,i)$'s $\tcpa$ constraint if $\cC(1,i)$ can not afford the full item. However, in this case, we can simply increase the fraction of the item $N(1,i)_i$ (or $H(p,s)$ respectively) that $\cC(1,i)$ gets and decrease the fraction that $\cD(1,i)$ (or $\cT_1$ respectively) gets in the allocation vector of the uniform-bidding equilibrium until $\cC(1,i)$'s $\tcpa$ is binding, and the result is still a uniform-bidding equilibrium. Thus,
\begin{align*}
&\textrm{bidder $\cC(1,s)$'s CPA}=\textrm{bidder $\cC(1,s)$'s $\tcpa$}=
\frac{n^3+n+\delta_1\sum_{t\in[n]}A_{st}+3/2}{2n^3+4n-2},\\
&\textrm{bidder $\cC(1,s')$'s CPA}=\textrm{bidder $\cC(1,s')$'s $\tcpa$}=
\frac{n^3+n+\delta_1\sum_{t\in[n]}A_{s't}+3/2}{2n^3+4n-2}.
\end{align*}
It follows that
\begin{equation}\label{eq:CPA_diff}
    \textrm{bidder $\cC(1,s)$'s CPA}-\textrm{bidder $\cC(1,s')$'s CPA}=\frac{\sum_{t\in[n]}\delta_1A_{st}-\sum_{t\in[n]}\delta_1A_{s't}}{2n^3+4n-2}.
\end{equation}

Next, we calculate bidder $\cC(1,s)$'s and bidder $\cC(1,s')$'s total payment and total value respectively to get different bounds for their CPAs.

First, for any $i\in[n]$ (including $s$ and $s'$), since except $\cC(1,i)$, only bidder $\cC(1,t)$ bids $\tilde{x}_t$ on the item $N(1,i)_t$ ($\tilde{x}_t$ is $\cC(1,t)$'s bidding parameter, and $1$ is $\cC(1,t)$'s value for $N(1,i)_t$), and by Lemma~\ref{lem:C_pacing_param_le_3}, $\tilde{x}_t<3$ (which is less than bidder $\cC(1,i)$'s bid $3\tilde{x}_i\ge3$), it follows that bidder $\cC(1,i)$ wins all the items $\{N(1,i)_t\mid t\in[n]\textrm{ and } t\neq i\}$ and pays $\sum_{t\in[n]\textrm{ and } t\neq i}\tilde{x}_t$. Moreover, by Lemma~\ref{lem:D_gets_le_delta_2}, bidder $\cC(1,i)$ wins at least $1-2\delta_2$ fraction of $N(1,i)_i$ (and at most the full item), and because bidders $\cC(1,i)$ and $\cD(1,i)$ have the same bidding parameter by Lemma~\ref{lem:matching_pacing_param} and the same value $1$ for $N(1,i)_i$, $\cC(1,i)$ pays at least $\tilde{x}_i(1-2\delta_2)$ (and at most $\tilde{x}_i$) for $N(1,i)_i$. Moreover, since except $\cC(1,i)$, only bidder $\cC(2,t)$ bids $\tilde{y}_t\delta_1A_{it}$ on the item $E(1,i)_t$ ($\tilde{y}_t$ is $\cC(2,t)$'s bidding parameter, and $\delta_1A_{it}$ is $\cC(2,t)$'s value for $E(1,i)_t$), and by Lemma~\ref{lem:C_pacing_param_le_3}, $\tilde{y}_t\delta_1A_{it}<3\delta_1A_{it}$ (which is less than bidder $\cC(1,i)$'s bid $\tilde{x}_i\ge1$ for $E(1,i)_t$), it follows that bidder $\cC(1,i)$ wins all the items $\{E(1,i)_t\mid t\in[n]\}$ and pays $\sum_{t\in[n]}\delta_1A_{it}\tilde{y}_t$. Furthermore, bidder $\cC(1,i)$ wins the item $L(1,i)$ for free since there is no competition for this item. In addition, it is easy to verify that with bidding parameter $<3$, bidder $\cC(1,i)$ can not get any fraction of the items $\{E(2,t)_i\mid t\in[n]\}$ and $\{N(1,t)_i\mid t\in[n]\textrm{ and } t\neq i\}$.

Finally, we calculate $\cC(1,i)$'s value and cost for the item $H(1,i)$. To this end, we need to do case analysis for $s$ and $s'$ (because although bidder $\cC(1,s)$'s bidding parameter $\tilde{x}_s>1$ by our assumption, bidder $\cC(1,s')$'s bidding parameter $\tilde{x}_{s'}$ could be $>1$ or exactly $1$). Since $\cT_1$ is the only bidder other than $\cC(1,s)$ bids $n^3$ on the item $H(1,s)$ ($n^3$ is $\cT_1$'s value for $H(1,s)$, and $1$ is $\cT_1$'s bidding parameter by Lemma~\ref{lem:T1_pacing_param_is_1}), it follows that $\cC(1,s)$ wins $H(1,s)$ by paying $n^3$. Similarly, $\cT_1$ also bids $n^3$ on the item $H(1,s')$. However, since $\tilde{x}_{s'}$ can be $>1$ or exactly $1$, we only know that bidder $\cC(1,s')$ wins at least a fraction of the item $H(1,s')$. Let $\tau$ denote the fraction of $H(1,s')$ which $\cC(1,s')$ wins, and then $\cC(1,s')$ pays $\tau n^3$ for this item.

In summary, bidder $\cC(1,s)$ gets total value at most $2n^3+4n-2$ and makes total payment at least $n^3+\sum_{t\in[n]}\tilde{x}_t-2\delta_2\tilde{x}_{s}+\sum_{t\in[n]}\delta_1A_{st}\tilde{y}_t$. Thus,
\begin{align*}
    \textrm{bidder $\cC(1,s)$'s CPA}&\ge\frac{n^3+\sum_{t\in[n]}\tilde{x}_t-2\delta_2\tilde{x}_{s}+\sum_{t\in[n]}\delta_1A_{st}\tilde{y}_t}{2n^3+4n-2}\\
    &=\frac{n^3+\sum_{t\in[n]}\tilde{x}_t+\sum_{t\in[n]}\delta_1A_{st}\tilde{y}_t-O(\delta_2)}{2n^3+4n-2}.
\end{align*}
Bidder $\cC(1,s')$ gets total value at least $(1+\tau)n^3+4n-2-2\delta_2$ and makes total payment at most $(1+\tau)n^3+\sum_{t\in[n]}\tilde{x}_t+\sum_{t\in[n]}\delta_1A_{s't}\tilde{y}_t$. Thus,
\begin{align*}
    \textrm{bidder $\cC(1,s')$'s CPA}\le&\frac{\tau n^3+\sum_{t\in[n]}\tilde{x}_t+\sum_{t\in[n]}\delta_1A_{s't}\tilde{y}_t}{(1+\tau) n^3+4n-2-2\delta_2}\\
    \le& \frac{n^3+\sum_{t\in[n]}\tilde{x}_t+\sum_{t\in[n]}\delta_1A_{s't}\tilde{y}_t}{2n^3+4n-2-2\delta_2}\\
    &\text{(because the first upper bound is maximized when $\tau=1$)}\\
    =& \frac{n^3+\sum_{t\in[n]}\tilde{x}_t+\sum_{t\in[n]}\delta_1A_{s't}\tilde{y}_t}{2n^3+4n-2}\cdot\left(1+\frac{2\delta_2}{2n^3+4n-2-2\delta_2}\right)\\
    =& \frac{n^3+\sum_{t\in[n]}\tilde{x}_t+\sum_{t\in[n]}\delta_1A_{s't}\tilde{y}_t+O(\delta_2)}{2n^3+4n-2}.
\end{align*}
Combining our lower bound of bidder $\cC(1,s)$'s CPA and upper bound of bidder $\cC(1,s')$'s CPA, we get
\begin{equation}\label{eq:CPA_diff_lower_bound}
    \textrm{bidder $\cC(1,s)$'s CPA}-\textrm{bidder $\cC(1,s')$'s CPA}\ge \frac{\sum_{t\in[n]}\delta_1A_{st}\tilde{y}_t-\sum_{t\in[n]}\delta_1A_{s't}\tilde{y}_t-O(\delta_2)}{2n^3+4n-2}.
\end{equation}
Putting Eq.~\eqref{eq:CPA_diff} and Eq.~\eqref{eq:CPA_diff_lower_bound} together, we have that
\[
    \sum_{t\in[n]}\delta_1A_{st}\tilde{y}_t-\sum_{t\in[n]}\delta_1A_{s't}\tilde{y}_t-O(\delta_2)\le\sum_{t\in[n]}\delta_1A_{st}-\sum_{t\in[n]}\delta_1A_{s't},
\]
which implies that
\[
    \sum_{t\in[n]}A_{st}(\tilde{y}_t-1)-\sum_{t\in[n]}A_{s't}(\tilde{y}_t-1)\le O(\delta_2/\delta_1)=O(1/n^2).
\]
This is exactly Eq.~\eqref{eq:approximate_best_response}, which completes the proof.
\end{proof}

\section{Proof of Upper Bound of Price of Anarchy}\label{app:sec5}
\begin{proof}[Proof of Theorem~\ref{lem:tight-PoA}]
We will construct two instances such that $PoA=O(1/\log(T_{max}/T_{min}))$ for the first instance, and $PoA=O(1/\log(\beta_{max}/\beta_{min}))$ for the second instance. The theorem follows by picking the instance with worse $PoA$ upper bound from those two instances (depending on which of $T_{max}/T_{min}$ and $\beta_{max}/\beta_{min}$ is larger). We let $\varepsilon=\frac{1}{8^n}$.
\medskip

\noindent {\em The $\tcpa$-instance.} \textbf{Bidders:} There is a $\tcpa$ bidder $j_2$ with $\tcpa$ $1$, and there is another set of $\tcpa$ bidders $J_1:=\bigcup_{\ell\in[n]} J_1^{\ell}$, where $J_1^{\ell}$ contains $2^{n-\ell}$ $\tcpa$ bidders with $\tcpa$ $2^{\ell}$ for each $\ell\in[n]$. Moreover, there is two $\ql$ bidders $q_1$ and $q_2$.

\textbf{Channels:}
There are two channels $k_1$ and $k_2$. Channel $k_2$ owns only one impression $i_2$ which is of value $1$ to bidder $j_2$ and value zero to everyone else. Channel $k_1$ owns impressions $\{i_1,i_1'\}\cup\{i_{1}^{j}\mid j\in J_1\}$. Impression $i_1$ is of value $1-2\varepsilon$ to bidder $j_2$, value $1$ to bidder $q_1$, and value zero to everyone else. Impression $i_1'$ is of value $\varepsilon$ to bidder $j_2$, value $1$ to bidder $q_2$, and value zero to everyone else. For any $j\in J_1^{\ell}$ for each $\ell\in[n]$, impression $i_{1}^{j}$ is of value $1$ to bidder $j$, value  $\varepsilon 2^{\ell}$ to bidder $j_2$, and value zero to everyone else.

\textbf{Global model:} We first consider the global model and prove that the total revenue is $n 2^n$ if both channels set zero reserve prices. Since only bidder $j_2$ has non-zero value for impression $i_2$, and the reserve price is zero, bidder $j_2$ will get $i_2$ of value $1$ for zero cost. Thus, bidder $j_2$'s $\tcpa$ constraint is not tight if $j_2$ only gets impression $i_2$. It follows by item (4) of Definition~\ref{def:uniform_bidding_equilibrium} that bidder $j_2$ should increase the bidding parameter until getting tied for a new impression.

Now we show that bidder $j_2$ must be tied first for impression $i_1$ and then impression $i_1'$, before getting tied for any impression in $\{i_{1}^{j}\mid j\in J_1\}$. Notice that by Assumption~\ref{lem:tcpa-bid-atleast-tcpa}, bidder $q_1$ would bid $1$ for impression $i_1$, and bidder $q_2$ would bid $1$ for impression $i_1'$, and bidder $j\in J_1^{\ell}$ for any $\ell\in[n]$ would bid at least $2^{\ell}$ for impression $i_1^j$. For bidder $j_2$ to be tied with a bid $1$ for impression $i_1$, $j_2$'s bidding parameter only needs to be $\frac{1}{1-2\varepsilon}$, and moreover, for bidder $j_2$ to be tied with a bid $1$ for impression $i_1'$, $j_2$'s bidding parameter needs to be $\frac{1}{\varepsilon}$, and furthermore, for bidder $j_2$ to be tied with a bid $2^{\ell}$ for impression $i_{1}^{j}$ with $j\in J_1^{\ell}$ for any $\ell\in[n]$, $j_2$'s bidding parameter needs to be at least $\frac{2}{\varepsilon}$. Thus, $j_2$ must be tied for impression $i_1$ first. Notice that even if impression $i_1$ is fully sold to bidder $j_2$ at a cost $1$, bidder $j_2$'s total spend for impressions $i_1$ and $i_2$ divided by their total value is $\frac{1}{2-2\varepsilon}$, which is still below $j_2$'s $\tcpa$. It follows by item (4) of Definition~\ref{def:uniform_bidding_equilibrium} that bidder $j_2$ will increase the bidding parameter to $\frac{1}{\varepsilon}$ to be tied for impression $i_1'$ and get a small fraction of $i_1'$.

From the discussion of the above two paragraphs, it follows that bidder $j_2$'s bidding parameter is at least $\frac{1}{\varepsilon}$. Hence, bidder $j_2$'s bid is at least $2^{\ell}$ for impression $i_{1}^{j}$ with $j\in J_1^{\ell}$ for any $\ell\in[n]$, and because $j_2$ is bidding above the reserve price for impression $i_{1}^{j}$ (which is zero), by item (2) of Definition~\ref{def:uniform_bidding_equilibrium}, impression $i_{1}^{j}$ must be fully sold (for a cost at least $2^{\ell}$). It follows that channel $k_2$'s revenue is at least $\sum_{\ell\in[n]}2^{\ell}\cdot |J_1^{\ell}|=\sum_{\ell\in[n]}2^{\ell}\cdot 2^{n-\ell}=n2^n$.

\textbf{Local model:} By Assumption~\ref{lem:tcpa-bid-atleast-tcpa}, bidder $j_2$ would use a bidding parameter at least $j_2$'s $\tcpa$ $1$. Thus, in the local model, if channel $k_2$ sets a reserve price $1-\varepsilon$ for impression $i_2$, by item (2) of Definition~\ref{def:uniform_bidding_equilibrium}, $i_2$ will be fully sold to bidder $j_2$ for a cost $1-\varepsilon$ in the subgame equilibrium regardless of channel $k_1$'s reserve price. Hence, we know that channel $k_2$'s revenue in the local model is at least $1-\varepsilon$. In other word, in the local model, bidder $j_2$ spends at least $1-\varepsilon$ for impression $i_2$.

Now we show that in the local model, bidder $j_2$'s bidding parameter is at most $\frac{1}{1-2\varepsilon}$ in the subgame equilibrium. Suppose for contradiction bidder $j_2$'s bidding parameter is strictly larger than $\frac{1}{1-2\varepsilon}$, then it follows that $j_2$'s bid for impression $i_1$ is strictly larger than $1$, and thus, bidder $j_2$ would win the full impression $i_1$ for a cost larger than $1$. Therefore, the total spend for impressions $i_1$ and $i_2$ divided by their total value is at least $\frac{2-\varepsilon}{2-2\varepsilon}>1$ (and taking other impressions into account will only make this worse as other impressions need bidder $j_2$ to use even higher bidding parameter), which violates bidder $j_2$'s $\tcpa$ constraint.

Therefore, with a bidding parameter at most $\frac{1}{1-2\varepsilon}$, bidder $j_2$ bids at most $\frac{\varepsilon2^{\ell}}{1-2\varepsilon}$ for impression $i_{1}^{j}$ with $j\in J_1^{\ell}$ for any $\ell\in[n]$. Note that $\frac{\varepsilon2^{\ell}}{1-2\varepsilon}\le\frac{\varepsilon2^{n}}{1-2\varepsilon}\le 2^{-n}$ by our choice of $\varepsilon$. Thus, bidder $j_2$'s bid for impression $i_{1}^{j}$ with $j\in J_1^{\ell}$ for any $\ell\in[n]$ is negligible compared to the value of impression $i_{1}^{j}$ to bidders $j$, which means that bidder $j_2$'s bid can only make a negligible difference for the cost of impression $i_{1}^{j}$ to bidders $j$. Finally, since channel $k_2$ uses a uniform reserve price, and $J_1$ is essentially the same ``equal-revenue'' instance as in Lemma~\ref{ex:revg-ub}, we have that the revenue of channel $k_2$ is $O(2^n)$.

To conclude, $PoA$ for our $\tcpa$ instance is $O(1/n)=O(1/\log(T_{max}/T_{min}))$.

\medskip
\noindent {\em The $\budgeted$-instance.}
The construction is analogous to the $\tcpa$-instance:

\textbf{Bidders:} There is a $\budgeted$ bidder $j_2$ with budget $1$, and there is another set of $\budgeted$ bidders $J_1:=\bigcup_{\ell\in[n]} J_1^{\ell}$, where $J_1^{\ell}$ contains $2^{n-\ell}$ $\budgeted$ bidders with budget $2^{\ell}$ for each $\ell\in[n]$. Moreover, there is two $\ql$ bidders $q_1$ and $q_2$.

\textbf{Channels:}
There are two channels $k_1$ and $k_2$. Channel $k_2$ owns only one impression $i_2$ which is of value $1$ to bidder $j_2$ and value zero to everyone else. Channel $k_1$ owns impressions $\{i_1,i_1'\}\cup\{i_{1}^{j}\mid j\in J_1\}$. Impression $i_1$ is of value $1$ to bidder $j_2$, value $1-\varepsilon$ to bidder $q_1$, and value zero to everyone else. Impression $i_1'$ is of value $\varepsilon$ to bidder $j_2$, value $1$ to bidder $q_2$, and value zero to everyone else. For any $j\in J_1^{\ell}$ for each $\ell\in[n]$, impression $i_{1}^{j}$ is of value $1$ to bidder $j$, value  $\varepsilon 2^{\ell}$ to bidder $j_2$, and value zero to everyone else.

The proof of $PoA \leq O(1/\log({\beta_{max}}/{\beta_{min}}))$ for the $\budgeted$-instance is analogous to our proof for the $\tcpa$ instance.
\end{proof}

\section{Proofs for Scaled Channels}\label{ap:sec-publisher}
\subsection{Proof of Theorem~\ref{th: scaled channels PoA=0}}
 \begin{proof}[Proof of Theorem~\ref{th: scaled channels PoA=0}]
 Consider the following instance. There are two symmetric channels (i.e., $\gamma_k = 1/2)$ and 
 two $\tcpa$ bidders with targets $T_1=2$ and $T_2=1$. There are five types of impressions owned by the channels ordered by the publisher price on each of them: $I_1, I_2, I_3, I_4$ and $I_5$ where $\delta>\epsilon$. The publishers reserve prices and bidders valuations are described in Figure~\ref{fig:1}.
 
\begin{figure}[h!]
  \centering
    \includegraphics[width=0.9\textwidth]{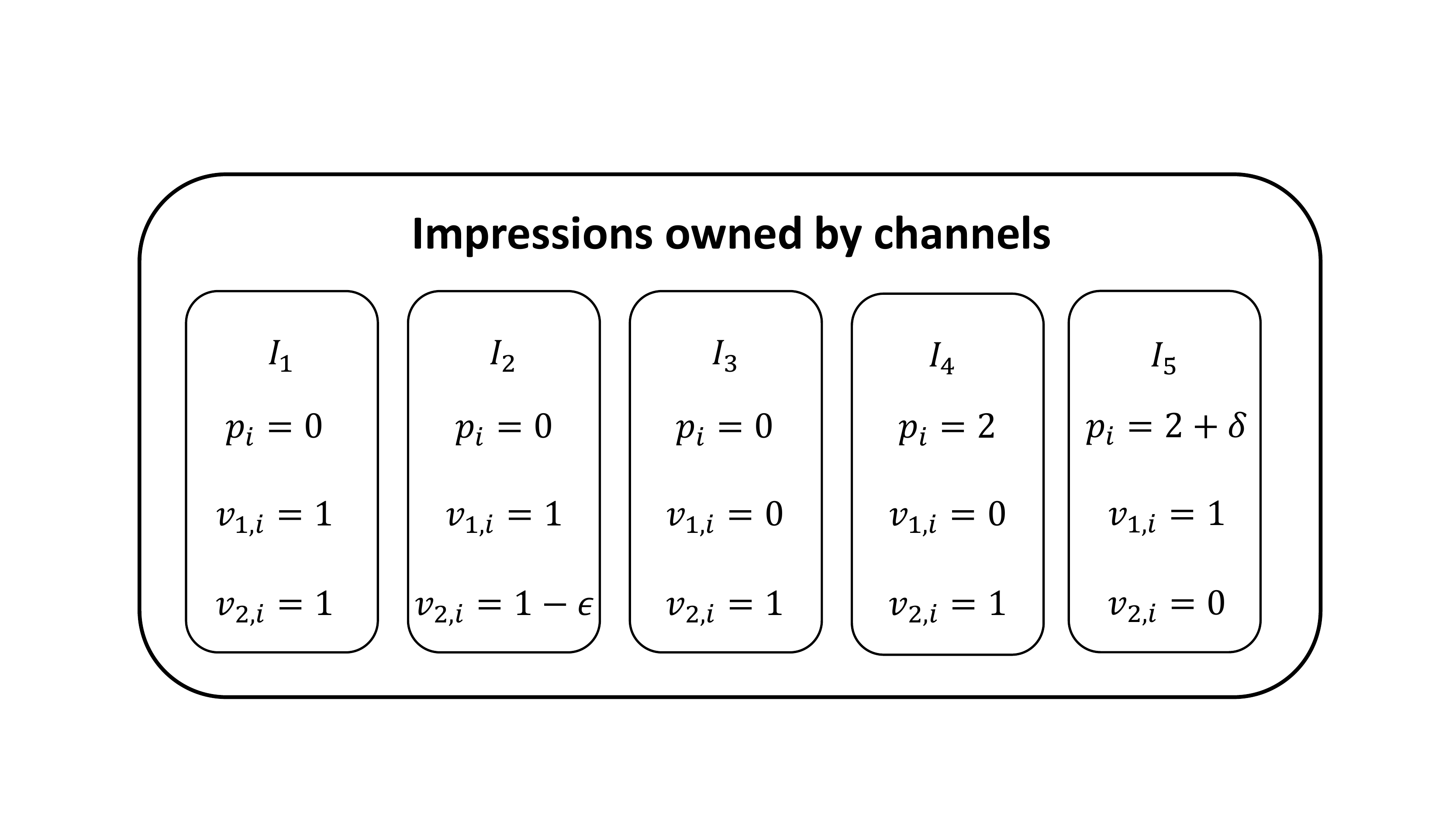}
      \caption{Instances used in Theorem~\ref{th: scaled channels PoA=0} to show that $PoA=0$.}\label{fig:1}
\end{figure}

\noindent{\bf Global model:} Observe that for this case a feasible solution for the channels is to set a uniform reserve price of $1$. For this subgame, since all impressions cost at least $1$, Bidder 2 does not have any slack to buy impressions that are more than $1$. Hence, Bidder 2 bids $b_{2,i} = v_{2,i}$. Consequently, Bidder $1$ bids $b_{1,i} = (2+\delta)v_{1,i}$. The outcome of the auctions is that Bidder 1 gets all impressions $I_1,I_2$ for a price of $1$ and a subset of impressions of $I'_5\subseteq I_5$ for a price $2+\delta$ with $|I'_5| =(|I_1|+(1+\epsilon)|I_2|)/\delta$ so that Bidder 1 $\tcpa$ constraint is tight. Thus, in this subgame, the revenue collected by the two channels is approximately $ |I_1|+|I_2|+|I_3|+ (1+2/\delta)(|I_1|+|I_2|)$ for small $\epsilon$. Since setting a reserve price of $1$ in both channels is a feasible policy in global model, we conclude that for small $\epsilon$, $Rev(Global)\geq |I_1|+|I_2|+|I_3|+ (1+2/\delta)(|I_1|+|I_2|)$.
\medskip

\noindent{ \bf Local model:} We assert that there is an equilibrium where both channels set reserve prices to $0$. In the on-path subgame of the equilibrium, Bidder $1$ bids $b_{1,i} =  (2+\delta)v_{1,i}$ and Bidder $2$ bids $b_{2,i}= 2v_{2,i}$.  With this bidding strategies, Bidder 1 gets all impressions $I_1,I_2$ and, by assuming $|I_1|<|I_2|$, we have that Bidder $1$ has slack to get a subset $I_5'\subseteq I_5$ to make its $\tcpa$ constraint binding. Bidder $2$ gets all impressions $I_3$ and subset $I'_4\subseteq I_4$ to make its $\tcpa$ constraint binding. Under this bidding behavior, for small $\epsilon$, we have that $I'_4,I'
_5$ are small $O(\epsilon)$. Hence, the revenue each channel obtains is approximately $(2(|I_1|+|I_2|) + |I_3|)/2$. 

We now show that setting reserve prices to $0$ is an equilibrium for the channels. Since the game is symmetric for each channel we only consider channel 1's deviations. If Channel 1 sets a reserve $r_1> 2+\delta$, in the subgame, Bidder 1 and Bidder 2 only buy impressions from Channel 2. Hence, Channel 1 gets a revenue of $0$. If $r_1\in (2, 2+\delta]$, Bidder $2$ buys impressions only from channel 2 which implies that Channel 1 loses $|I_3|/2$ relative to setting $r_1=0$. Bidder 1, instead, pays $r_1$ for impressions $I_1,I_2$. Compared to setting a reserve $r_1=0$, the gain channel 1 obtains from bidder 1 is no more than $\delta (|I_1|+|I_2|)$. Thus, for small $\delta$, deviating to $r_1\in [2+\epsilon, 2+\delta)$ is not profitable. If $r_1\leq 2$, then the revenue coming from Bidder $1$ is the same as the case of $r_1=0$, since the price is determined by bidder $2$'s bid. Regarding the revenue coming from bidder $2$, we have that Channel 1 gets $(1-\epsilon)r_1$ on impressions $I_3$ and gets $x$ impressions of $I_4$ where $x= \frac{(1-\epsilon) (2-r_1)}{2(p_4-1)}|I_3|$. Thus, the revenue gains by Channel 1 is $(1-\epsilon)|I_3| + x (1-p_4/2)$. Since $p_4=2$ we have that Channel 1 is indifferent on setting any reserve price $r_1\in [0,2]$. We conclude that setting a reserve price $r_1=0$ is optimal and hence an equilibrium. 

To conclude the proof, by comparing the global and local models we obtain that for $\epsilon$ small,   
$$ PoA \leq \frac{2(|I_1|+|I_2|) + |I_3|} {|I_1|+|I_2|+|I_3|+ (1+2/\delta)(|I_1|+|I_2|)}.$$
We conclude the proof by taking $\delta \to 0$.
\end{proof}

\subsection{Proof of Theorem~\ref{poa: 1/k}}

We split the proof Theorem~\ref{poa: 1/k} in the following steps.

 

First, we show that to bound the $PoA$ without loss of generality we can focus on instances where the bidder is a $\tcpa$ bidder.

\begin{lemma}\label{lem:60}
Suppose that there is a single bidder in the game. If the bidder is either a $\budgeted$ bidder or $\ql$ bidder then $RevG(Local)=RevG(Global)$.
\end{lemma}
\begin{proof}
If the bidder is a $\ql$ bidder, the channel's reserve price optimization problem is independent of the other channels. Thus, both Global and Local models achieve the same revenue. 

If the bidder is a $\budgeted$ bidder, we claim that in all equilibria of the local model the bidder spend its budget. Suppose not, then one of the channel can slightly increase the reserve price while keeping the $\budgeted$ bidder unconstrained. Thus, the bidder does not change its bids and the channel deviating increases its revenue, which is contradiction. We conclude that the total revenue in local model is the bidder's budget which matches the optimal liquid welfare. This implies that $RevG(Local)=RevG(Global)$.
\end{proof}

The next step characterizes the optimal reserve price for the global model.  

\begin{lemma}[Global model: optimal reserves]\label{lem:s6:1}
For a single $\tcpa$ bidder with constraint $T$ the solution of the global model is that every channel sets the lowest reserve price such that the $\tcpa$ bidder constraint is tight. That is, channels set $\underline r = \argmin \{ 
\sum_{i \in I} v_i = T \sum_{i\in I}\max\{r,p_i\} \}$.
\end{lemma}

\begin{proof}[Proof of Lemma~\ref{lem:s6:1}]

The proof-idea follows from the fact that the revenue from a $\tcpa$-constraint bidder is roughly $\tcpa$-target times the volume of impressions acquired by the bidder. Because volume is inversely proportional to reserve prices, in the global model, all channels set a reserve price as low as possible conditional that the $\tcpa$ constraint remains binding.

We fist show that in the global model without loss all channels can set the same reserve price. Indeed, because the bidder is using a uniform bid across all impressions, we have that its final bid in the value-space is the same in each channel. In particular, observe that from the bidder's standpoint it is equivalent to face a reserve price $r_k$ on Channel $k$'s impressions or to face the same reserve price $r= \sum_{k \in K} \gamma_k r_k$ in all channels. Thus, from the bidder's perspective its bidding behavior does not change by facing a symmetric reserve price across channels. Likewise, because the final bid remains unchanged the global revenue does not change by using the symmetric reserve prices. 

In what follows, we denote by $r$ such symmetric reserve price. 

Let $V(q)$ the bidder's value of impressions having publisher reserve price less than $q$. That is,  
\begin{align*}
    V(q) = & \max \sum_{i\in I} v_i x_i \\
    &\mbox{s.t. }\; \sum_{i\in I} p_i x_i \leq q,\\
   & \qquad  x_i \in [0,1].
\end{align*}
Observe that $V$ is a continuous function, and hence, can be uniformly approximated by differentiable functions over compact sets \citep{rudin}. Hence, by a simple limiting argument, we can assume that $V$ is differentiable.

Given a symmetric reserve price $r$ and a bid multiplier $\alpha\geq \max\{r,T\}$, the value the bidder obtains is $V(\alpha)$ for a cost of $ r V(r) + \int_r^\alpha qV'(q)dq= \alpha V(\alpha) - \int_{r}^\alpha V(q)dq$. 

Therefore, the bidder's best response when channels set a reserve price $r$ is given by the solution $\alpha(r)$ to equation
\begin{equation}\label{eq:V} (\alpha - T) V(\alpha) = \int_r^{\alpha} V(z) dz.
\end{equation}

We derive the following observation from $\alpha(r)$.

{\bf Observation 1}: $\alpha(r)$ is non-increasing as function of $r$. To see this, notice that the left-hand-side of Equation~\eqref{eq:V} is independent of $r$ while the right-hand-side of the equation is decreasing in $r$. Thus, we get that $\alpha(r)\leq \alpha(r')$ for $r'<r$.  

To conclude the proof, we define  
\begin{equation}\label{eq:small_r}
\underline r = \min \{r \mbox{ s.t.  Equation~\ref{eq:V} has solution}\} 
\end{equation}.\footnote{Notice that $\underline r$ is well-defined as the set of $r$ solving Equation~\ref{eq:V} is compact and non-empty (for $r=T, \alpha (T)=T$ solves the equation).} 
and claim that $\underline r$ is the optimal reserve price. Indeed, for $r>\underline r$ we have that $\alpha(r)< \alpha(\underline r)$ due to Observation 1. Hence, the total revenue with $r$ is $T\cdot V(\alpha(r))<T\cdot V(\alpha(\underline r))$. 

For $r<\underline r$, notice that Equation~\eqref{eq:V} does not have solution. This means that the bidder is buying all impression without making the $\tcpa$ constraint binding. In other words, the bidder is buying the same impressions but for a cheaper price. We conclude that $\underline r$ is the optimal reserve price.
\end{proof}

 The following lemma characterizes the bidding equilibrium of the largest channel in the local model.

\begin{lemma}[Local model: large channel reserve]\label{lem:l6:final}
Consider a single $\tcpa$ bidder with constraint $T$ and a (pure strategy) reserve prices equilibrium satisfying that $\sum_{k \in K}\gamma_k r_k > \underline r$ for $\underline r$ defined in Equation~\eqref{eq:small_r}. When channels are not symmetric (i.e. $\gamma_k \neq \gamma_{k'}$ for some $k, k'$), every {large} channel $\k$ ($\gamma_{\k}\geq \gamma_{k'}$ for $k'\neq k$) sets a reserve price $r_{\k} = \underline r$.
\end{lemma}

This lemma shows that in the local model, the competition among channels leads to larger channels setting efficient reserve prices and providing value to the bidder, improving the efficiency of the allocation. In turn, small channels raise their reserve price extracting the value provided from larger channels and creating revenue inefficiencies when aggregating all channels.


\begin{proof}[Proof of Lemma~\ref{lem:l6:final}]
Fix an equilibrium for the channels $(r_k)_{k\in K}$ such that $\sum_{k \in K}\gamma_k r_k > \underline v$.





Consider a local deviation where one channel decreases the reserve price. Let $s$ the extra-value the bidder obtains when channel $k$ lowers their reserve price. The bidder will spend this extra-value on more impressions. That is, in the subgame, the bidder reacts to the decrease in reserve price by increasing its bid from $\alpha$ to $\alpha_s$, satisfying that
\begin{equation}\label{eq:surp}
    \int_{\alpha}^{\alpha_s}(q-T)V'(q)dq = s
\end{equation} 
where $V$ is defined in Equation~\eqref{eq:V} in Lemma~\ref{lem:s6:1}. 

Under this deviation, the revenue gain by channel $k$ is $\gamma_k \int_{\alpha}^{\alpha_s}q V'(q)dq$ while it exerts a cost of $s$. From equilibrium condition we have that $\gamma_k \int_{\alpha}^{\alpha_s}q V'(q)dq\leq s$. Taking $s\to 0$ (i.e. taking the limit of the deviation on $r_k$ to zero), we get that $ \gamma_k \alpha V'(\alpha) d\alpha_s/ds - 1 \leq 0$. From Equation~\eqref{eq:surp}, we also obtain that $(\alpha - T)V'(\alpha)d\alpha_s/ds=1$. Plugging these two expressions, and noticing that $V'(\alpha)>0$ and $d\alpha_s/ds<0$,\footnote{$\alpha_s$ is decreasing on $s$ by the same argument used for Observation 1 in Lemma~\ref{lem:s6:1}.} we conclude that a channel does not benefit by locally reducing its reserve price if and only if $T\geq \alpha (1-\gamma_k)$. 

Conversely, using the same argument we conclude that a channel does not benefit by increasing its reserve price if and only if $T\leq \alpha (1-\gamma_k)$. 

To conclude the proof, suppose for the sake of a contradiction that $r_{\k} > \underline v$ for one of the largest channel $\k$. Then, it is feasible for such channel to reduce their reserve price. Thus, in equilibrium we must have that $T\geq \alpha (1-\gamma_{\k})$. Because channels are not symmetric, there is a channel $k'$ with $\gamma_{k'}< \gamma_{\k} $.  Then, for channel $k'$ we have that   $T> \alpha (1-\gamma_{k'})$. This implies that it is channel $k'$ would increase their revenue by increase their reserve price from $r_{k'}$ to $r_{k'} + \epsilon$, for some small $\epsilon$. This contradicts the equilibrium assumption. 
\end{proof}

After the preliminaries steps we are now in position to proof Theorem~\ref{poa: 1/k}.

\begin{proof}[Proof of Theorem~\ref{poa: 1/k}]

From Lemma~\ref{lem:60}, we can restrict our attention to the case where the bidder is a $\tcpa$ bidder.  \medskip

{\bf Proof that $PoA\geq 1/k$}\medskip
 
 Consider $(r_k)_{k\in K}$ an arbitrary pure-strategy reserve price equilibrium on the local model. There can be the following 3 possibilities.

{\bf Case 1.} That $\sum_{k \in K}\gamma_k r_k < \underline v $. This case cannot be an equilibrium: it implies that the bidder is unconstrained. Hence, one channel can slightly increase its reserve price while keeping the bidder unconstrained, and hence, keeping the same the bid. Therefore, a channel would increase its revenue which is a contradiction.

{\bf Case 2.}  If $\sum_{k \in K}\gamma_k r_k = \underline v$, then the bidder faces the same reserve as in the global optimal solution. Thus, for that case, the outcome of the local model is the same as of the global model.

{\bf Case 3.} When $\sum_{k \in K}\gamma_k r_k > \underline v$. If channels are asymmetric (i.e., $\gamma_k \neq \gamma_{k'}$ for some $k, k'$), Lemma~\ref{lem:l6:final} implies that the largest channel sets a reserve $r_{\k}= \underline v$. Now, a feasible solution for the $\tcpa$ bidder is to purchase only impressions from one of the largest channel. By doing this suboptimal bidding strategy, the bidder gets a total value of $\gamma_{\k}k V(\alpha(\underline r))$. Thus, in an equilibrium, the bidding parameter $\alpha_{LocalEQ}$ must satisfy $V(\alpha_{LocalEQ}) \geq \gamma_k H (\alpha (\underline r))$. Since the bidder is $\tcpa$-constrained, the total revenue across the channels is the target $T$ times the value the bidder gets in an equilibrium. Using the reserve in the global model is $\underline r$ (Lemma~\ref{lem:s6:1}), we conclude that 
$$PoA  = \inf_{\alpha_{LocalEQ}} \frac {T \cdot V(\alpha_{LocalEQ})}{T\cdot V(\alpha (\underline r))} \geq  \frac {\gamma_k  V(\alpha (\underline r))}{V(\alpha (\underline r))} = \gamma_k\geq \frac 1 k.$$

It remains to tackle Case 3. when channels are symmetric, i.e., $\gamma_k = 1/k$. Using the same argument for deviations as in Lemma~\ref{lem:l6:final}, we have that either one channel sets reserve $r= \underline r$ or all channels are setting the same reserve price in which case $r$ is such that $T = \alpha_{LocalEQ}(1-1/k)$. If one channel sets $r= \underline r$ the same proof as the asymmetric case holds. If not, since the optimal bid multiplier also satisfy Equation~\eqref{eq:V}, we get that
\begin{align*}
(\alpha_{Global}-T) V(\alpha_{Global}) &= \int_{\underline r}^{\alpha_{Global}} V(z)dz \\
&=\int_{\underline r}^{r} V(z)dz + \int_{ r}^{\alpha_{LocalEQ}} V(z)dz + \int_{\alpha_{LocalEQ}}^{\alpha_{Global}} V(z)dz \\
& = \int_{\underline r}^{r} V(z)dz + (\alpha_{LocalEQ}-T) V(\alpha_{LocalEQ}) + \int_{\alpha_{LocalEQ}}^{\alpha_{Global}} V(z)dz \\
&\leq  (r-\underline r) V(\alpha_{LocalEQ}) + (\alpha_{LocalEQ}-T) V(\alpha_{LocalEQ}) \\
&\quad + (\alpha_{Global} -\alpha_{LocalEQ})  V(\alpha_{Global}).
\end{align*}
Where the last inequality holds since $V$ is non-decreasing and $\alpha_{LocalEQ}\leq \alpha_{Global}$.

Rearranging terms and noticing that $(r-\underline r) V(\alpha_{LocalEQ}) + (\alpha_{LocalEQ}-T) \leq \alpha_{LocalEQ} V(\alpha_{LocalEQ})$ holds since $\underline r\leq r\leq T$, we obtain that 
$$ (\alpha_{LocalEQ} -T) V(\alpha_{Global}) \leq \alpha_{LocalEQ} V(\alpha_{LocalEQ}).$$
We conclude that $PoA\geq 1/k$ by using the equilibrium condition we have that $T = \alpha_{LocalEQ}(1-1/k)$. This implies that $V(\alpha_{LocalEQ})/V(\alpha_{Global}) \geq 1/k$.
\medskip

{\bf Proof that $PoA\leq 1/k$}
\medskip

To proof the tightness of the $PoA$, we consider an instance with $k$ symmetric channels ($\alpha_k =\frac 1 k)$. There are two types of impressions the high impressions $H$, and the low impressions $L$ that each channel owns. The publisher pricing constraint for the high impressions is $p_i=h$ for $i\in H$ and for the low impressions is $p_i=l$ for $i\in L$. 
We consider that there are $|H|=n_1$ high impressions and $|L|=n_2$ low impressions. The $\tcpa$ constraint is $T=1$. 

We assume that $n_1(h-1)=n_2$. This implies that  $\underline r = 0$ and that $Rev_{Global} = T\cdot (h+l)$ (Lemma~\ref{lem:s6:1}).

We assert that in the local model, there is an equilibrium where all channels set a reserve price $r_k=1$. To see this, suppose a Channel deviates to $r'_k$ and assume that $r'_k$ is an optimal deviation. Then, we have the following cases to study.
\begin{itemize}
\item If $r'_k>1$, since the $\tcpa$ bidder does not have slack the bidder does not buy from channel $k$. This is not a profitable deviation.
\item If $r'_k<1$ and the $\tcpa$ bidder is not able to buy some of the high impressions $H$, then the deviation is unprofitable: without the deviation, the channel is selling impressions $L$ at a price $1>r'_k$. 
\item If $r'_k<1$ and the $\tcpa$ bidder is able to buy some impressions in $H$.  Then if $h$ is such that $1< h (1-1/k)$,  we have that the channel can further improve its revenue by slightly increasing its reserve price higher than $r'_k$ (the proof for the condition on the profitable deviation is in Lemma~\ref{lem:l6:final}). This contradicts the optimality of $r'_k$. 
\end{itemize}
We conclude that so long as $1< h (1-1/k)$, setting reserve price $r_k=1$ is an equilibrium for the local game. 

In this equilibrium of the local model, the total revenue across channels is $n_2$ 

Therefore, 
\begin{align*}
    PoA &\leq \frac {n_2}{n_2+ n_1}
    = \frac{n_1 (h-1)}{n_1(h-1)+n_1}
    = 1- \frac 1 h.
\end{align*}
By taking the limit when $h \to \frac 1 {1-1/k}$, we conclude that $PoA\leq 1/k$.
\end{proof}

\section{Sketch of results for Welfare}
\label{app:welfare}
Most of our revenue and Price of Anarchy results carry over to {\em welfare}.\footnote{For welfare, we define the Price of Anarchy of the Local model vs. Global model as $PoA = \inf \frac{Wel(x_{Local})}{Wel(x_{Global})}$.} In particular for the setting without publisher reserves,
\begin{enumerate}
    \item The revenue lower bound in Theorem~\ref{thm:rev-local} carries over easily, as revenue is a lower bound on welfare, and the benchmark we use is the optimal Liquid Welfare.
    \item The example for the revenue upper bound in Proposition~\ref{ex:revg-ub} can be modified to get a similar upper bound on welfare. This can be done by adding a few extra bidders -- a couple with TCPA $1$ and a couple with budget $1$.
    \item A similar modification of the example in Theorem~\ref{lem:ineq-POA} gives us an upper bound on Price of Anarchy for welfare.
    \item Putting these together, we get a tight bound (up to constants) on Price of Anarchy for welfare, similar to Theorem~\ref{thm:poa-main}.
\end{enumerate}
The proofs are deferred to the full paper.

\bigskip

\end{document}